\documentclass[aps,10pt,prd,notitlepage,showpacs,nofootinbib,superscriptaddress]{revtex4-1}
\usepackage{graphicx}

\usepackage{amsmath}
\usepackage{amssymb}
\usepackage{siunitx}
\usepackage[section]{placeins}
\usepackage{float}
\usepackage{comment}
\usepackage{slashed}
\usepackage[normalem]{ulem}
\usepackage{braket}
\usepackage{comment}

\usepackage{caption}
\usepackage{subcaption}

\usepackage[usenames,dvipsnames]{color}
\usepackage[colorinlistoftodos]{todonotes}
\usepackage[colorlinks=true,citecolor=darkred,urlcolor=darkred, pdfborder={0 0 0}]{hyperref}
\definecolor{darkred}{rgb}{0.6,0,0}
 
\newcommand {\ignore}[1]{}

\newcommand{\bea}{\begin{eqnarray}}
\newcommand{\eea}{\end{eqnarray}}

\def\gsim{\raise0.3ex\hbox{$\;>$\kern-0.75em\raise-1.1ex\hbox{$\sim\;$}}}
\def\lsim{\raise0.3ex\hbox{$\;<$\kern-0.75em\raise-1.1ex\hbox{$\sim\;$}}}

\usepackage{soul}

\definecolor{mightnightblue}{RGB}{25,25,112}

\definecolor{brown}{rgb}{0.59, 0.29, 0.0}

\def\21{$\mathrm{SU(2)_L \otimes U(1)_Y}$}

\begin{document}
\bibliographystyle{unsrt}   


\title{Probing Electroweak Phase Transition in Extended Singlet Scalar Model with Resonant $HH$ production in $bbZZ$ Channel using Parameterized Machine Learning}

\author{Pritam Palit}
\email{pritampalit@gmail.com}
\affiliation{Saha Institute of Nuclear Physics, 1/AF Bidhannagar, Kolkata 700064, India}
\affiliation{Homi Bhabha National Institute, Training School Complex, Anushakti Nagar, Mumbai 400085, India} 

\author{Sujay Shil}
\email{sujayshil1@gmail.com}

\affiliation{Instituto de F\'isica, Universidade de S\~ao Paulo, R. do Mat\~ao 1371, 05508-090 S\~ao Paulo, Brazil}

\date{\today}


\begin{abstract}
In this paper, a collider signature of a heavy Higgs boson at $14$ TeV HL-LHC is studied, where the heavy Higgs boson decays into a pair of standard model Higgs boson, which further decays to $bbZZ$ state and subsequently to $bb\ell^{+} \ell^{-}\nu_{\ell} \nu_{\ell}$ final state. To study this, we consider singlet scalar extension of the standard model and select the parameter space and mass of the heavy Higgs boson such that it prefers a strong first-order electroweak phase transition. The study is done following the $bbZZ$ analysis of CMS Collaboration and further using parameterized machine learning for final discrimination which simplifies the training process along with an improved discrimination between signal and background over the range of benchmark points. Despite the lower branching fraction, this channel can be a potential probe of the electroweak phase transition with the data sets collected by the CMS and ATLAS experiments at the $14$ TeV HL-LHC with $3$ $\rm{ab}^{-1}$ of integrated luminosity and a production of resonant di-Higgs signal can be potentially discovered up to 490 GeV of resonance mass.
\end{abstract}
\maketitle
\section{Introduction}
Discovery of the Higgs boson at Large Hadron Collider \cite{CMS:2012qbp, ATLAS:2012yve} in 2012 completes the discovery of all the fundamental particles predicted in the standard model (SM). However,  there are many questions which remain unanswered, for example, origin of matter-anti matter asymmetry \cite{Planck:2013pxb}, mechanism of neutrino mass generation \cite{AKHMEDOV2000215}, nature of dark matter \cite{1933AcHPh...6..110Z, 1937ApJ....86..217Z} etc. 
The mechanism by which matter anti-matter asymmetry is generated, is called baryogenesis. Depending on the characteristics of electroweak phase transition (EWPT), one can account how much baryon asymmetry is created by standard model and beyond standard model (BSM) processes. To have successful electroweak baryogenesis, three Sakharov  conditions \cite{Sakharov:1967dj} should be satisfied which includes the deviation from thermal equilibrium or CPT symmetry needs to be broken as imposed by the third Sakharov condition. Considering only the standard model Higgs doublet, electroweak phase transition is not a strong first-order phase transition. Rather it is a crossover transition \cite{Aoki:1999fi, Csikor:1998eu, Laine:1998jb, Gurtler:1997hr} and it can not generate sufficient baryon asymmetry of the universe. However, some variations of the standard model for example, singlet scalar extension of the standard model, denoted by the xSM, actually can make electroweak phase transition a strong first-order electroweak phase transition (SFOEWPT) \cite{Profumo:2007wc, Espinosa:2011ax, Ghorbani:2018yfr}. In this paper, we mostly analyze the collider signature of the xSM considering the parameter space which satisfy the SFOEWPT. 
\paragraph*{} Currently, CMS and ATLAS have performed resonant di-Higgs searches \cite{CMS:2022dwd, ATLAS:2022vkf} in different decay final states : $4b$ \cite{ATLAS:2018rnh, CMS:2018qmt}, $bbWW$ and $bbZZ$ \cite{ATLAS:2018fpd, CMS:2017rpp, PhysRevD.102.032003},  $bb \tau \tau$ \cite{ATLAS:2018uni, CMS:2017hea}, $bb \gamma \gamma$ \cite{ATLAS:2018dpp, CMS:2018tla}, $WWWW$ \cite{ATLAS:2018ili} and $\gamma \gamma WW$ \cite{ATLAS:2018hqk}. So far, there has been no significant excess over SM backgrounds. On the other hand, there are a number of phenomenological studies on the parameter regions suitable for SFOEWPT. 
A discovery up to 500 GeV of heavy singlet-like Higgs mass is possible in the $bb \gamma \gamma$, $4 \tau$ final states \cite{Kotwal:2016tex} and $4b$ \cite{Li:2019tfd} final state with a luminosity of 3 $\rm{ab}^{-1}$ at the 14 TeV high-luminosity LHC (HL-LHC). In the $bbWW$ final state, a discovery is achievable in the range of [350, 600] GeV at the 13 TeV LHC with a luminosity of 3 $\rm{ab}^{-1}$ \cite{Huang:2017jws}. There are also existing studies of di-Higgs discovery reach for HL-LHC using other models like generic two-Higgs-doublet models (2HDM) in different final states \cite{Lu:2015qqa, Ren:2017jbg}. 
 \paragraph*{}In this study, using the xSM, we analyze the prospects of exclusion or discovery in the $bbZZ$ channel at the 14 TeV HL-LHC with a luminosity of 3 $\rm{ab}^{-1}$. We consider $bb \ell \ell\nu\nu (\ell=e, \mu)$ decay of $bbZZ$, as this has not been covered yet by other studies and there is
an existing study by CMS \cite{CMS-PAS-HIG-17-032}  which is used to validate the analysis methodology. Few points are considered to use this final state instead of $bb 4\ell$ or $bb 2j2\ell$. For the $bb 4\ell$ channel, there is no existing 13 TeV results from CMS or ATLAS for resonant di-Higgs production, so that we can validate our results.
Between $bb 2j2\ell$ and $bb \ell \ell\nu\nu$, the first one has larger branching ratio, where the second one does not suffer from multijet backgrounds.
Also, we already have an existing study on the same final state of bbWW \cite{Huang:2017jws}, which acts as a relevant comparison with that of bbZZ.
In principle, $bb 4\ell$ or $bb 2j2\ell$ channels can also be studied in details, but this is beyond the scope of this paper.
 We assume a final combination of both CMS and ATLAS experiments. We choose the benchmark points with mass of BSM Higgs having a range between 300 GeV and 850 GeV producing the maximum signal rate of resonant di-Higgs production and satisfying all the recent experimental constraints coming from electroweak precision data and the Higgs signal rate as well as the theoretical constraints from perturbativity, vacuum stability and a SFOEWPT. 
 \paragraph*{} There are several phenomenological studies of resonant di-Higgs production using multivariate analysis and machine learning \cite{Alves:2018oct, Alves:2019emf}. In our analysis, once the basic event selection is done, for final signal-to-background discrimination we use parameterized machine learning \cite{Baldi:2016fzo, Anzalone_2022} method to apply a single deep neural network (DNN) algorithm for all the benchmark points taking the mass of the resonant particle as a parameter to the network. This parameterized DNN replaces the individual classifiers trained at individual mass points of the resonant particle with a single classifier and simplifies the training process with improved performance. Finally, the signal significance is obtained by the parameterized DNN score distributions of signal and background events. From our study and results shown in Fig. \ref{fig:comparesig}, we can conclude as follows : 
 \begin{itemize}
     \item Despite the lower branching ratio of $bbZZ$ channel, for singlet-like Higgs masses below 490 GeV, the combination of CMS and ATLAS experiments gives significance higher than $5\sigma$ in the $bbZZ$ final state along with the $bb\gamma \gamma$ and $4b$ channels and a discovery can be achieved at the 14 TeV HL-LHC with a luminosity of 3 $\rm{ab}^{-1}$ 
     for the regions of parameter space where SFOEWPT is satisfied.
     \item Only CMS search does not achieve discovery significance in this channel, but we can exclude up to 550 GeV of singlet-like Higgs masses.
 \end{itemize}
\paragraph*{} As per the previous studies done in Ref. \cite{Alves:2018oct, Alves:2019igs, Alves:2020bpi}, our analysis can also be used for a complementary study of gravitational waves along with the di-Higgs production in collider search.
\paragraph*{}Our work is detailed in following ways : In Sec. \ref{theory}, we describe the overview of the xSM model along with the theoretical and phenomenological constraints, the requirements for a SFOEWPT and the benchmark points by scanning the parameter space. In Sec. \ref{13tev}, the validation of backgrounds is discussed by comparing with 13 TeV CMS analysis
results \cite{CMS-PAS-HIG-17-032}. In Sec. \ref{14tev}, we present the analysis of the signal and backgrounds at 14 TeV HL-LHC. Sec. \ref{summary} offers a summary and conclusion. In the Appendix \ref{apndx1}, the distributions of the kinematic variables are discussed which are used as input variables of the parameterized DNN for 14 TeV HL-LHC.
 

\section{The \MakeLowercase{x}SM Model \& EWPT}
\label{theory} 
 
 \subsection{The Model }
 \label{model}
 Most generalised potential of the scalar sector of the xSM takes the form~\cite{Profumo:2007wc, Espinosa:2011ax},
\begin{eqnarray}
\label{ScalarPotential1}
& V(H,S) =  -\mu^2 \left( H^\dagger H \right) + \lambda \left( H^\dagger H \right)^2 + \frac{\rm{a}_1}{2} \left( H^\dagger H \right) S & \nonumber \\
&  + \frac{\rm{a}_2}{2} \left( H^\dagger H \right) S^2 + \frac{\rm{b}_2}{2} S^2 + \frac{\rm{b_3}}{3} S^3 + \frac{\rm{b}_4}{4} S^4 , &
\end{eqnarray}
where $S$ is the real singlet scalar extension of the standard model. The coefficient $a_{1}$, $a_{2}$ are the couplings through which the SM Higgs doublet $H$ mixes with $S$. If $a_{1}$ and $b_{3}$ are absent, the potential is $\mathcal{Z}_{2}$ symmetric under the transformation $S\rightarrow -S$. However, we keep all the terms here, as these terms have significant role in electroweak phase transition as well as di-Higgs collider phenomenology.  \\
\paragraph*{}When the electroweak symmetry breaks, the Higgs doublet $H\rightarrow \frac{h+v_0}{\sqrt{2}}$ gets its vacuum expectation value (vev) and let assume the singlet scalar $S$ also gets a positive vev ($S\rightarrow x_0+s$). For the stability of the Higgs potential, the quartic Higgs coupling should be positive along all the directions, which gives $\lambda > 0$, $b_{4} > 0$ and $a_{2} > 2\sqrt{\lambda b_{4}}$. Once these conditions are satisfied, the Higgs potential definitely acquires one or multiple minima. The minimization conditions give the following sets of equations,
\begin{eqnarray}
\mu^{2}=\lambda v_{0}^{2}+(a_{1}+a_{2}x_{0})\frac{x_{0}}{2} ,  \\
b_{2} = -b_{3}x_{0}-b_{4}x_{0}^{2}-\frac{a_{1}v_{0}^{2}}{4x_{0}}-\frac{a_{2}v_{0}^{2}}{2} .
\end{eqnarray}
Also the condition for stable minimum gives,
\begin{eqnarray}
 b_3 x_0 + 2 b_4 x_0^2 - \frac{a_1 v_0^2}{4 x_0} - \frac{ (a_1 + 2 a_2 x_0 )^2 }{ 8 \lambda } > 0 .
\end{eqnarray}
In addition to that, the requirements for global minimum and perturbativity bound \cite{Robens:2015gla, Robens:2016xkb,  Gonderinger:2009jp} are imposed numerically, which are as follows, 
\begin{equation}
    \left \lvert \dfrac{a_1}{2} \right \rvert, \left \lvert \dfrac{a_2}{2} \right \rvert, \left \lvert \dfrac{b_4}{4}\right \rvert < 4\pi .
\end{equation}
After EWSB, the fields $h$ and $s$ mix with each other and the mass squared matrix elements are given by,
\begin{eqnarray}
& m_{h}^2 \equiv  \frac{d^2 V}{dh^2} = 2 \lambda v_0^2 & \nonumber \\
& m_{s}^2 \equiv  \frac{d^2 V}{ds^2}  = b_3 x_0 + 2 b_4 x_0^2 - \frac{a_1 v_0^2}{4 x_0} & \nonumber \\
 & m_{hs}^2 \equiv  \frac{d^2 V}{dh ds} = \left(a_1 + 2 a_2 x_0 \right) \frac{v_0}{2} .
 \label{mixingM}
 \end{eqnarray}
By diagonalizing the mass matrix one can get the mass eigenvalues as,
\begin{eqnarray}
& m_{1,2}^2 =  \frac{ m_{h}^2 + m_{s}^2 \mp \left| m_{h}^2 - m_{s}^2 \right| \sqrt{ 1 +  \left( \frac{ m_{hs}^2 }{ m_{h}^2 - m_{s}^2 } \right)^2 } } {2} . &
\label{Meigenvalues}
\end{eqnarray}
The corresponding mass eigenstates are,
\begin{eqnarray}
& h_1 = h \cos \theta + s \sin \theta , & \nonumber \\
& h_2 = - h \sin \theta + s \cos \theta , &
\label{eigstates}
\end{eqnarray}
where $\theta$ is the mixing angle between $h$ and $s$, which is defined as,
\begin{eqnarray}
\sin 2\theta=\frac{(a_{1}+2a_{2}x_{0})v_{0}}{m_{1}^{2}-m_{2}^{2}} .
\end{eqnarray}
Due to the mixing between the $h$ and $s$, 
standard model Higgs boson and BSM Higgs coupling with pairs of fermions or vector bosons (represented by $XX$)
are re-scaled by $\cos \theta$ and $\sin \theta$ respectively,
\begin{eqnarray}
g_{h_1 XX} = \cos \theta \; g_{h XX}^{\mathrm{SM}}\quad ,\quad \quad \quad g_{h_2 XX} = \sin \theta \; g_{h XX}^{\mathrm{SM}}.
\end{eqnarray}
%
To study the resonant di-Higgs production the most important interaction is $\lambda_{211}h_2\,h_1\,h_1$, where $\lambda_{211}$  is the tri-linear coupling given by,
\begin{eqnarray}
\label{g211}
\lambda_{211} &=& \frac{1}{4}\left[ (a_1 + 2 a_2 x_0)\, \cos ^3 \theta + 4 v_0 (a_2 -3 \lambda)\, \cos ^2 \theta \sin {\theta} \right. \nonumber\\
 &-& \left.2(a_1 + 2 a_2 x_0 -2 b_3 - 6b_4 x_0)\, \cos_\theta \sin ^2 \theta -2 a_2 v_0 \, \sin ^3 \theta \right] .
\end{eqnarray}
In this work, we only consider the mass hierarchy $m_{2} > 2 m_{1}$, so that an on-shell $h_{2}$ can decay into an $h_{1}$ pair. The decay width of $h_{2}$ to $h_{1}$ pair is given by,
%
%
\begin{eqnarray}
\Gamma_{h_{2} \rightarrow h_{1} h_{1}} = \frac{\lambda_{211}^2 \sqrt{1- 4m_{1}^2/m_{2}^{2}}}{8\pi m_2} ,
\end{eqnarray}
and the total width of $h_2$ is given by,
\begin{eqnarray}
\label{h2totalwidth}
 \Gamma_{h_2} &=& \sin^2\theta~\Gamma^{SM}(m_2) + \Gamma_{h_2 \rightarrow h_1 h_1} ,
\end{eqnarray}
where $\Gamma^{SM}(m_2)$ is the total width of the SM Higgs boson if its mass becomes $m_2$ and values of $\Gamma^{SM}(m_2)$ are taken from Ref. \cite{deFlorian:2227475}.
For $pp\rightarrow h_2\rightarrow XX$, the signal rate  normalized to the SM value is estimated by,
\begin{eqnarray}
\label{sigrate}
\mu_{h_2 \rightarrow X X} &=& \sin^2\theta \left( \frac{\sin^2\theta \Gamma^{SM}(m_2) }{\Gamma_{h_2}} \right) .
\end{eqnarray}
 For $pp\rightarrow h_2\rightarrow h_1 h_1$, the production cross-section can be written as,
\begin{eqnarray}
\label{sigxsec}
\sigma_{h_1 h_1} = \sigma^{SM}(m_2) \times \sin^2\theta \frac{\Gamma_{h_2 \rightarrow h_1 h_1}}{\sin^2\theta \Gamma^{SM}(m_2) + \Gamma_{h_2 \rightarrow h_1 h_1}} ,
\end{eqnarray}
where $\sigma^{SM}(m_2)$ is the LHC production cross section of SM Higgs boson when the
SM Higgs mass is equal to $m_2$.
\subsection{Existing Phenomenological Constraints}

\paragraph{} The Higgs signal strength measurements severely constrain the mixing angle of the Higgs boson with the BSM Higgs. A global fit is already performed using LHC run $2$ data by marginalizing up to $95\%$ confidence limit (C.L.) of the Higgs signal strength measurement values in Higgs production channels ($ggH$, $VBF$, $VH$ and $ttH$) and decays ($h\rightarrow \gamma\gamma$, $h\rightarrow\tau\tau$, $h_{1}\rightarrow WW$, $h_{1}\rightarrow ZZ$ and $h_{1}\rightarrow b b$) \cite{ATLAS:2018hxb, ATLAS:2018ynr, ATLAS:2018xbv, ATLAS:2017azn, ATLAS:2018kot,   ATLAS:2018mme}, and obtained a upper limit on $\sin^{2}\theta$, which is $0.131$ \cite{Li:2019tfd}. We also verify this limit. 
\paragraph{} There have been many LHC analyses which are focused on the searching of the heavy Higgs boson. We consider the bounds from the existing searches such as $h_{2}\rightarrow VV$ \cite{CMS:2018amk, ATLAS:2018sbw, ATLAS:2015pre,  ATLAS:2015oxt, CMS:2013vyt, CMS:2015hra} and also $h_{2}\rightarrow h_{1}h_{1}$, where $h_{1}h_{1}$ decays into $4b$ \cite{ATLAS:2018rnh, CMS:2018qmt}, $bb\gamma\gamma$ \cite{ATLAS:2018dpp, CMS:2018tla} and $bb\tau\tau$ \cite{ATLAS:2018uni, CMS:2017hea}. 
All these constraints have been considered and exclusion region in the plane of (${cos\theta, m_{2}}$) is shown in Ref. \cite{Huang:2017jws}. The benchmark points which we are going to consider here are chosen considering all of the above constraints. 
\paragraph{} The parameters of the xSM model are also constrained by the measurement of the electroweak precision observables (EWPO). The mixing between the SM Higgs and the BSM scalar deviates the values of the oblique parameters ($S$, $T$ and $U$) from standard model predictions \cite{CMS:2015hra, Hagiwara:1994pw, Baak:2014ora}, which constrain the $\sin\theta$ depending on the mass of the heavy BSM scalar. Considering presently measured values of the oblique parameters and deviations within $95\%$ C.L., the obtained upper limit on $\sin^{2}\theta$ is $0.12$ for $m_2=250$ GeV to $0.04$ for $m_2=950$ GeV  which also matches with the previous study \cite{Li:2019tfd}.

\subsection{EWPT constraints}
 \label{EWPT_constraints}
 To account for the present baryon asymmetry of the universe, the SM Higgs mass should be below $70$ GeV considering only one Higgs doublet in the theory. LHC already discovered Higgs boson with mass $125$ GeV, which disfavours SFOEWPT. However, SFOEWPT can be achieved by extending the standard model and in this study we consider singlet scalar extension of the standard model. The finite temperature effective potential is considered where $T=0$ tree level potential and gauge independent thermal mass correction to the potential are included so that electroweak symmetry at high temperature is restored \cite{Quiros:1994dr, Profumo:2014opa}. Here, $a_{1}$ and $b_{3}$ parameters create barrier between broken and unbroken phase which allows SFOEWPT. The term with co-efficient $a_2$ strengthens the first-order transition. At high temperature Ref. \cite{Profumo:2014opa, Pietroni:1992in} are followed and the $T$ dependent and gauge independent vevs can be written in a cylindrical coordinate representation as,
\begin{equation}
\frac{\overline{v}(T)}{\sqrt{2}} = \overline{\phi}(T)\cos\alpha(T) ~~~~  \overline{x}(T) = \overline{\phi}(T)\sin\alpha(T) , 
\end{equation}
where $\overline{v}(T=0)=v_{0}$ and $\overline{x}(T=0)=x_{0}$. To define the critical temperature we use the condition where the broken and unbroken phases are degenerate ($V^{T\neq 0}_{eff}(\phi,\alpha\neq\pi/2, T_c)=V^{T\neq 0}_{eff}(\phi,\alpha=\pi/2, T_c$). Using critical temperature we can calculate the quenching effect of the sphaleron process \cite{Morrissey:2012db}. For large quenching effect the first-order EWPT is strong, which requires,
\begin{equation}
\cos \alpha (T_c ) \,\frac{\bar{\phi} (T_c)}{T_c} \gtrsim 1.
\end{equation} 
We want to select the benchmark points which satisfy the SFOEWPT as well as all the bounds which are discussed in the previous section such as Higgs signal strength, LHC search for heavy Higgs, EWPO, etc. In order to do that, first the parameters $a_{1}$, $b_{3}$, $v_{0}$, $b_{4}$ and $\lambda$ are scanned in the following range,
 \begin{equation}
a_1/\text{TeV}, \quad b_3/\text{TeV} \in [-1, 1], \quad x_0/\text{TeV} \in [0, 1], \quad b_4, \lambda \in [0, 1].
\end{equation}  
 The rest of the parameters are calculated using the values of $v_{0}=246$ GeV and $m_{h}=125$ GeV. A naive perturbativity bound on the Higgs portal coupling $a_2/2 \lesssim 5$ is also applied. After scanning the above parameters and selecting the points which satisfy all the bounds of the previous section along with the perturbative bound, the qualified benchmark points are passed through $\rm{COSMOTRANSITION}$ package \cite{Wainwright:2011kj} to calculate the critical temperature, sphaleron energy, tunnelling rate from the unbroken phase to the broken phase etc., to check whether SFOEWPT conditions are satisfied. 
 
 \subsection{Identification of benchmark points}
 We choose the parameters which satisfy the requirements as discussed above. Depending on that, we select benchmark points having maximum signal rate 
 in 11 consecutive $h_2$ mass windows of equal width
 with a total range of $m_2 = [300, 850]$ GeV. Upper bound on $m_2$ at 850 GeV is decided by satisfying conditions of SFOEWPT. We do not consider benchmark points corresponding to minimum signal rate, as it is expected to have very low sensitivity, beyond the possible reach of 14 TeV HL-LHC for our final state. All the benchmark points are given in Table \ref{tab:bmpoint}. The benchmark points B3 and B4 are kept to compare with the previous studies \cite{Kotwal:2016tex, Li:2019tfd, Huang:2017jws} in spite of the $h_2 \rightarrow ZZ$ search \cite{CMS:2018amk} in CMS experiment already excludes those benchmark points.  

\begin{table}[]
     \centering
     \begin{tabular}{|c|c|c|c|c|c|c|c|c|c|c|c|c|c|c|}
     \hline
          Benchmark & $\cos \theta$ & $m_2$ & $\Gamma_{h_2}$ & $x_0$ & $\lambda$ & $a_1$ & $a_2$ & $b_3$ & $b_4$ & $\lambda_{111}$ & $\lambda_{211}$ & $\sigma$ & BR & $\sigma\times BR$ \\
           & & (GeV) & (GeV) & (GeV) & & (GeV)& & (GeV) & & (GeV) & (GeV) & (pb) & & (pb) \\ 
     \hline 
            BM1 & 0.974 & 327 & 0.929 & 60.9 & 0.17 & -490 & 2.65 & -361 & 0.52 & 45 & 62.2 & 0.56 & 0.33 & 0.1848 \\
            BM2 & 0.980 & 362 & 1.15 & 59.6 & 0.17 & -568 & 3.26 & -397 & 0.78 & 44.4 & 76.4 & 0.48 & 0.40 & 0.192 \\
            BM3 & 0.983 & 415 & 1.59 & 54.6 & 0.17 & -642 & 3.80 & -214 & 0.16 & 44.9 & 82.5 & 0.36 & 0.33 & 0.1188 \\
            BM4 & 0.984 & 455 & 2.08 & 47.4 & 0.18 &  -707 & 4.63 & -607 & 0.85 & 46.7 & 93.5 & 0.26 & 0.31 & 0.0806 \\
            BM5 & 0.986 & 511 & 2.44 & 40.7 & 0.18 & -744 & 5.17 & -618 & 0.82 & 46.6 & 91.9 &  0.15 & 0.24 & 0.036 \\
            BM6 & 0.988 & 563 & 2.92 & 40.5 & 0.19 &  -844 & 5.85 & -151 & 0.08 & 47.1 & 104 &  0.087 & 0.23 & 0.02001 \\ 
            BM7 & 0.992 & 604 & 2.82 & 36.4 & 0.18 &  -898 & 7.36 & -424 & 0.28 & 45.6 & 119 & 0.045 & 0.30 & 0.0135 \\
            BM8 & 0.994 & 662 & 2.97 & 32.9 & 0.17 & -976 & 8.98 & -542 & 0.53 & 44.9 & 132 &  0.023 & 0.33 & 0.00759 \\
            BM9 & 0.993 & 714 & 3.27 & 29.2 & 0.18 & -941 & 8.28 & 497 & 0.38 & 44.7 & 112 & 0.017 & 0.20 & 0.0034 \\
            BM10 & 0.996 & 767 & 2.83 & 24.5 & 0.17 &  -920 & 9.87 & 575 & 0.41 & 42.2 & 114 & 0.008 & 0.22 & 0.00176 \\
            BM11 & 0.994 & 840 & 4.03 & 21.7 & 0.19 & -988 & 9.22 & 356 & 0.83 & 43.9 & 83.8 & 0.007 & 0.08 & 0.00056\\            
      \hline      
     \end{tabular}
     \caption{Values of the xSM parameters for each of the benchmark points for the maximum $\sigma_{h_2} \times \rm{BR}_{h_2 \rightarrow h_1 h_1} $ value at the 14 TeV HL-LHC.}
     \label{tab:bmpoint}
 \end{table}

 \section{Reproduction of existing results of 13 T\MakeLowercase{e}V LHC}
 \label{13tev}
 To cross-check with the existing LHC results, the CMS analysis in Ref. \cite{CMS-PAS-HIG-17-032} is followed and the distributions of final pseudo transverse mass (defined later in this section) of the backgrounds in the signal region are reproduced to justify the proper generation of backgrounds. In this CMS analysis, they look for the resonant search of graviton and radion for various range of masses in di-Higgs decay mode where the final state is $bbZZ\rightarrow bb\ell^{+} \ell^{-}\nu_{\ell} \nu_{\ell}$. 
 The top-pair ($t\bar{t}$) production as the dominant SM background and Drell-Yan with 2 jets (DY) as another sub-dominant background are simulated at leading order (LO) accuracy with MadGraph5\_aMC@NLOv2.6.7 \cite{Alwall:2014hca}. For simplicity, Monte Carlo generation of our signal and backgrounds is restricted to $l = \mu$ . When we further estimate the
sensitivity, the possible contributions from final states with a pair of electrons are taken into account, for which very similar efficiencies are expected.
\paragraph*{} PYTHIA8 is used \cite{Sjostrand:2014zea} for parton showering, fragmentation and hadronization. We use DELPHES3 \cite{deFavereau:2013fsa} for simulating the detector effects where we use the default CMS DELPHES card . 
The anti-$k_t$ clustering algorithm is used to reconstruct jets having a radius parameter R fixed at 0.4. The efficiency and misidentification probability of b-tagging are implemented by the CMVA algorithm \cite{CMS-PAS-BTV-15-001} inside the CMS Delphes card. In the analysis, medium working point of the CMVA algorithm is used which is defined such as the efficiency is about 66\%, while the mis-identification rate is about 1\% . 
 \\For the analysis, the selection criteria is used as follows :
 \begin{itemize}
     \item Events are required to have at least two muons both of which will be within $| \eta | < 2.4 $. The transverse momentum of the leading muon needs to be larger than 20 GeV, while for the subleading muon the minimum requirement is 15 GeV.
     \item Two jets constituting the $H \rightarrow bb$ candidate have to have $p_T > 30$ GeV and $| \eta | < 2.4 $ , and to be separated from leptons by a distance of $ \Delta R > 0.3$.
     \item One of the $Z$ bosons is reconstructed from the two highest $p_T$ muons. 
     \item Two b-jets are selected using CMVA algorithm. If the selection criteria is satisfied by more than two b-jets, we select the two b-jet candidates having a combined invariant mass closer to $m_{h_1} = 125 $ GeV.
     \item The reconstruction of the di-Higgs candidates is done in the regions that are chosen in the kinematic space defined by the dilepton invariant mass $m_{\ell \ell}$ and the invariant mass $m_{bb}$ of the pair of b-jets. The signal region (SR) is defined by the requirements $80 < m_{\ell \ell} < 96$ GeV to select only events with
on-shell Z bosons decaying into charged leptons and $ 105 < m_{bb} < 135$ GeV to select the Higgs boson decaying into the pair of b-jets.
     \item Missing transverse energy, $\rm{\slashed{E}}_{\rm{T}}$ is imposed to be larger than 25 GeV
     to suppress one of the major backgrounds, DY. 
     \item To differentiate signal and backgrounds in the SR, training of a Boosted Decision Tree (BDT) \cite{10.1214/aos/1013203451, Hocker:2007ht} is done on simulated MC samples that include a mix of the $t\bar{t}$ and DY samples representing the dominant backgrounds and a resonant signal sample as described below in details. Details of the BDT architecture used is given in Appendix \ref{apndx2}.
     \item  Pseudo transverse mass of the di-Higgs (HH) candidate is formed as, $\Tilde{M}_T (H H)= \sqrt{E^2 - p{_z}^2}$, where E and $p_z$ are denoted by the energy and the z-axis component of the Lorentz energy-momentum vector of the HH candidate, which is constructed as the
sum of the Lorentz vectors of the two leptons, two b-jets, and the four-vector ($\slashed{E}_T,\vec{p}_{T}^{miss}$), where $\slashed{E}_T$ is the magnitude of $\vec{p}_{T}^{miss}$, representing neutrinos as the z-component of the neutrinos' momentum is unknown.
Distributions of $\Tilde{M}_T (H H)$ are used as the final discriminant to match with the CMS analysis as in Ref. \cite{CMS-PAS-HIG-17-032}.
 \end{itemize} 

For BDT, 
the input variables include $\Delta{\rm{R}}$ between two b-jets, $\Delta{\rm{R}}$ between two leptons and the $\rm{\slashed{E}}_{\rm{T}}$, the invariant mass and the tranverse momentum of the on-shell Z boson and both Higgs boson candidates. The invariant mass of the Higgs boson that decays to two Z bosons ($m_{ZZ}$) is 
an approximation of its invariant mass obtained by summing up the four-vectors of the two charged leptons and the four-vector of the missing transverse momentum.
To compare with the CMS results, we train a BDT including the signal resonant di-Higgs samples with the mass of the resonance from 260 to 450 GeV and the combined background samples. The selection is applied according to BDT output distributions of signal sample for mass 300 GeV and combined background to achieve the maximum 
signal significance.

\begin{figure*}[h!]
    \centering
    \includegraphics[width=0.60\textwidth]{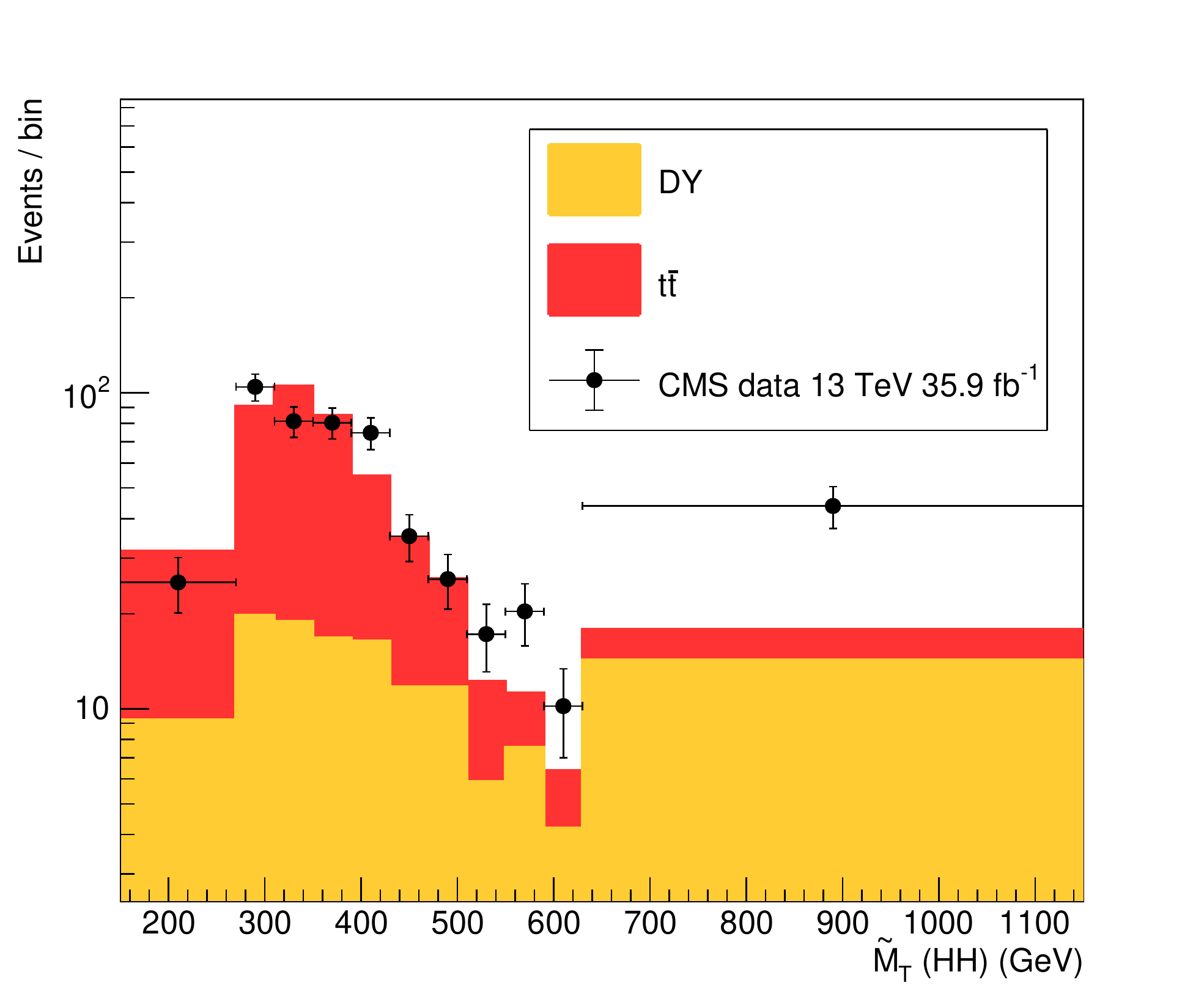}
    \caption{Distribution of the di-Higgs pseudo transverse mass in the signal region for variable bin widths. Data points are taken from the CMS data in Ref. \cite{CMS-PAS-HIG-17-032} and histograms represent the background processes from the simulation in the current analysis.}
     \label{fig:MT_13tev}
\end{figure*}

Fig. \ref{fig:MT_13tev} shows the di-Higgs pseudo transverse mass distributions of the background in the $\mu + \mu -$ channel for the graviton resonance mass hypothesis for resonance mass of 300 GeV. 
NLO K factor for DY is taken to be 1.135 \cite{Alwall_2014} while for $t\bar{t}$, NNLO + NNLL K-factor is taken to be 2.15 \cite{ttbarxsec}. We observe comparable agreement between the background estimate obtained by our analysis and CMS data points from Ref. \cite{CMS-PAS-HIG-17-032}, which validates the production of background samples. We observe a discrepancy in the higher pseudo transverse mass region. The CMS analysis considers Drell-Yan sample in association with up to four jets, while we consider Drell-Yan sample in association with up to two jets in our analysis. After the final SR selection, there are very low statistics left for the DY sample. To train the BDT, one needs to produce a large number of events at the initial stage for this background sample, which was beyond our scope for the production of the sample in association with up to four jets. The smaller number of jets in our Drell-Yan sample compared to the CMS analysis is the most likely cause for the discrepancy in the higher mass region of pseudo transverse mass given the fact that, due to the detector mismeasurements, the missing transverse momentum is positively correlated with the jet transverse momentum and multiplicity and the fact that the agreement is better in the CMS analysis.
Also, as there is negligible presence of signal statistics in this higher mass region of pseudo transverse mass, it justifies that this discrepancy will rarely affect our final estimate of significance for the 14 TeV HL-LHC prediction.

 \section{Projections for 14 T\MakeLowercase{e}V HL-LHC}
 \label{14tev}
 \subsection{Analysis strategy}
 Once the validation of our backgrounds against the CMS results at the 13 TeV LHC is done, the projections are estimated at the 14 TeV HL-LHC with a revised analysis strategy.
 The primary selections of events are kept similar to 13 TeV due to the similarity of distributions of kinematic variables. 
 \begin{itemize}
     \item Events are selected having at least two muons of opposite electric charge, both of which will be within $| \eta | < 2.4 $. The transverse momentum of leading muon requires to be larger than 20 GeV, while the subleading muon has the minimum requirement of 10 GeV.
     \item Two jets constituting the $H \rightarrow bb$ candidate have to have $p_T > 30$ GeV and $| \eta | < 2.4 $ , and to be separated from leptons by a distance of $ \Delta R > 0.3$.
     \item One of the $Z$ bosons is reconstructed from the two highest $p_T$ muons. 
     \item Two b-jets are selected using CMVA algorithm. If all selection criteria are satisfied by more than two b-jets, we select the two b-jet candidates having the highest $p_T$.
     \item The di-Higgs candidates are reconstructed in signal regions defined by the dilepton invariant mass $m_{\ell \ell}$ and the invariant mass $m_{bb}$ of the two b-jets. The signal region (SR) is defined by the requirements $80 < m_{\ell \ell} < 96$ GeV and $95 < m_{\rm{bb}} < 135$ GeV.
     \item For final discrimination between signal and background in the SR, a parameterized neural network \cite{Baldi:2016fzo} is trained on simulated MC samples that include a mix of the $t\bar{t}$ and DY samples representing the backgrounds and the resonant signal sample as described below in details.
 \end{itemize} 

\subsection{Parameterized neural network for final discrimination}
\begin{figure*}[h!]
    \centering
\begin{subfigure}{.495\textwidth}
    \includegraphics[width=0.85\linewidth]{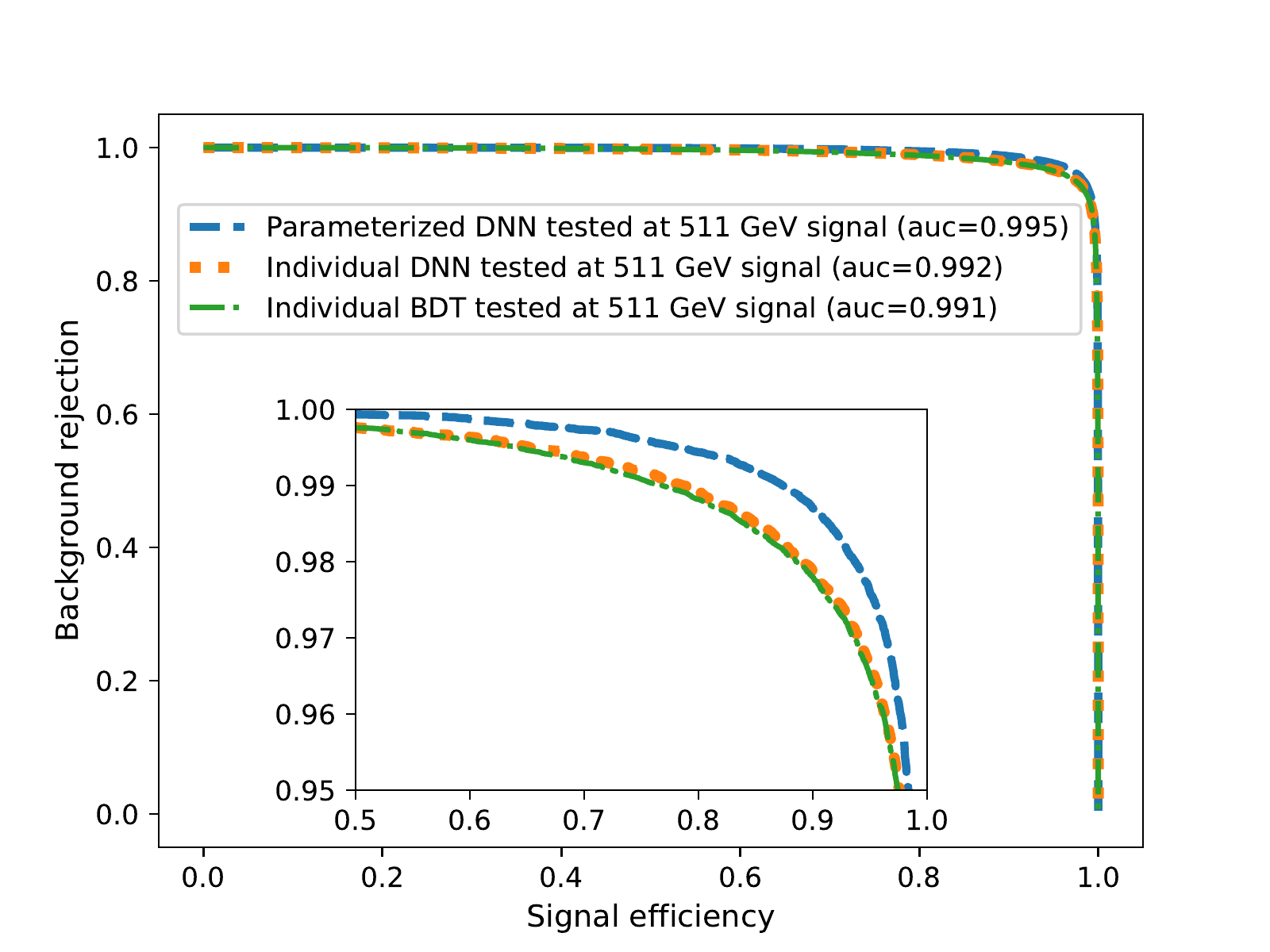}
    \caption{Performance comparison}
     \label{fig:paradnncompare}
\end{subfigure}
\hfill
\begin{subfigure}{.495\textwidth}
    \includegraphics[width=0.99\linewidth]{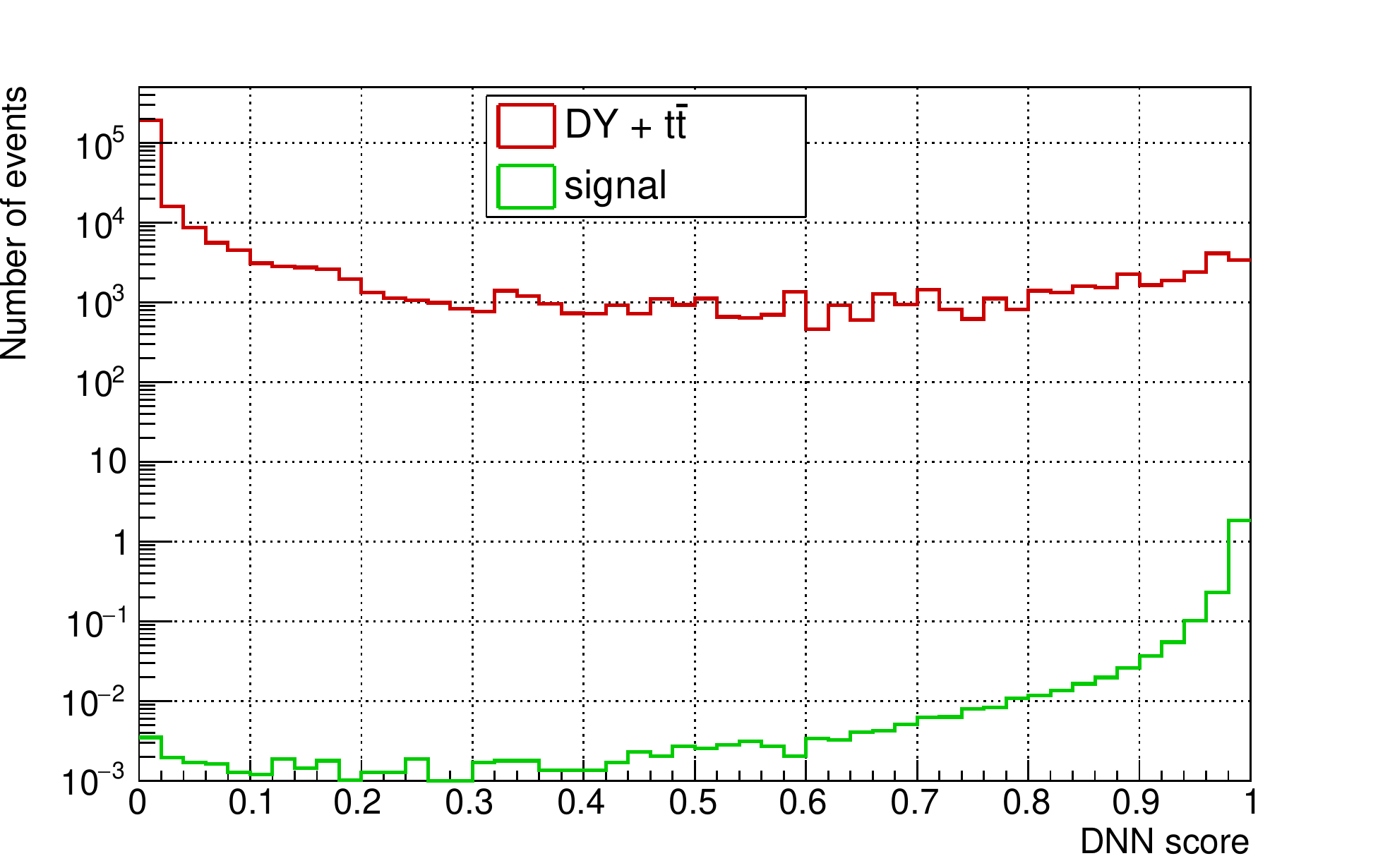}
    \caption{Parameterized DNN score}
     \label{fig:phenodnn}
\end{subfigure}
\caption{Here (a) shows the performance comparison of parameterized DNN as a ROC curve of background rejection ($1-\rm background\, efficiency$) vs signal efficiency at $511$ GeV signal mass corresponding to BM5 given in Table \ref{tab:bmpoint}. The blue dashed line, orange dotted line and green dashdot line show the parameterized DNN, individual DNN and BDT tested at the same benchmark point BM5, respectively. Values of the area under the curve (auc) metric are given to quantify the performance. (b) represents distributions of the parameterized DNN score at BM5. The green and red lines show the distributions for signal and background, respectively, normalized to $3~ \rm{ab^{-1}}$, where the background comprises of $t\bar{t}$ and DY samples. Log scale
for the y axis is used to show both shape and normalization differences.
}
\end{figure*}

Machine learning algorithms like boosted decision tree \cite{10.1214/aos/1013203451}, neural network \cite{Goodfellow-et-al-2016} are used frequently in signal-background classification in high energy physics experiments to achieve a higher sensitivity. Usually, a set of separate and independent classifiers is trained for each value of the input parameters (i.e. mass of the resonant particle in our case) corresponding to specific signal hypothesis. To separate the signal and background at each mass point, these individual classifiers perform much better compared to the classifier which is trained combining all of the masses together and evaluated on individual signal masses. But, one of the limitations of this approach is that a proper framework of training, testing and tuning including different hyperparameters needs to be maintained for each individual classifier, which makes the idea of having individual classifier at each individual signal mass point unfeasible.      
Parameterized neural network mitigates these issues by taking the true invariant mass of the resonant particle as a parameter to the network and adding this parameter as an additional input along with the other input features. The main advantage of the parameterized machine learning is the usage of a single network to evaluate the performance in different values of the parameter. Fig. \ref{fig:paradnncompare} shows the performance comparison among the parameterized DNN, individual DNN and individual BDT network. Parameterized DNN is trained combining all the signal mass points having the values of the true resonant masses corresponding to the mass points as input parameters and tested at 511 GeV signal mass corresponding to BM5. Detailed architecture of the parameterized DNN used is described in Appendix \ref{apndx2}. The individual DNN and BDT are trained and tested only at the 511 GeV mass point corresponding to BM5. It shows that with a single network, the parameterized DNN is able to perform as good as a network with a dedicated training for one individual signal mass point. The ROC curve for parameterized DNN gets comparatively higher value of area under the curve (auc) metric, as it is trained on higher number of events combining all the signal mass points, compared to individual DNN or BDT which used signal events of only one mass point for training. However, all the ROC curves have comparable performance, showing the effectiveness of applying parameterized DNN network. 
\paragraph*{} Parameterized DNN provides discrimination between signal and background across the range of invariant masses of resonant particle by a single network, which saves a lot of time as well as gives comparable performance compared to BDT or DNN trained on individual mass points. It also gives improved discrimination at intermediate mass points, as the network even learns the smooth interpolation in the signal masses between two signal mass points where the training of the network is not done. Detailed description of the set of variables used as input is given in Table \ref{table:vardnn}.
 The final parameterized DNN score distributions normalized to $3~ \rm{ab^{-1}}$ are shown in Fig. \ref{fig:phenodnn} for BM5.
The distributions of final output score of parameterized neural network are used for final discriminant of signal vs background to calculate the significance as a function of resonance mass following the profile likelihood ratio based $CL_S$ technique \cite{Zech:1988un}. 

\subsection{Final results and projection study}
The uncertainties for these backgrounds are considered here from the theoretical uncertainties coming from the factorization and renormalization scale variations and uncertainties in the parton distribution functions as well as the uncertainty on the integrated luminosity. For $t\bar{t}$, the factorization scale uncertainty and PDF uncertainties are taken directly from Ref. \cite{ttbarxsec}. For DY, the uncertainties from the theoretical cross-sections at 13 TeV \cite{Alwall:2014hca} is assumed to be reduced by a factor of 1/2 for 14 TeV HL-LHC \cite{CMS-PAS-FTR-18-019}. Besides, 1\% uncertainty on the total integrated luminosity is considered which affects both the yields of signal processes as well as background processes \cite{CMS-PAS-FTR-18-019}.

\begin{figure*}[h!]
    \centering
    \includegraphics[width=0.65\textwidth]{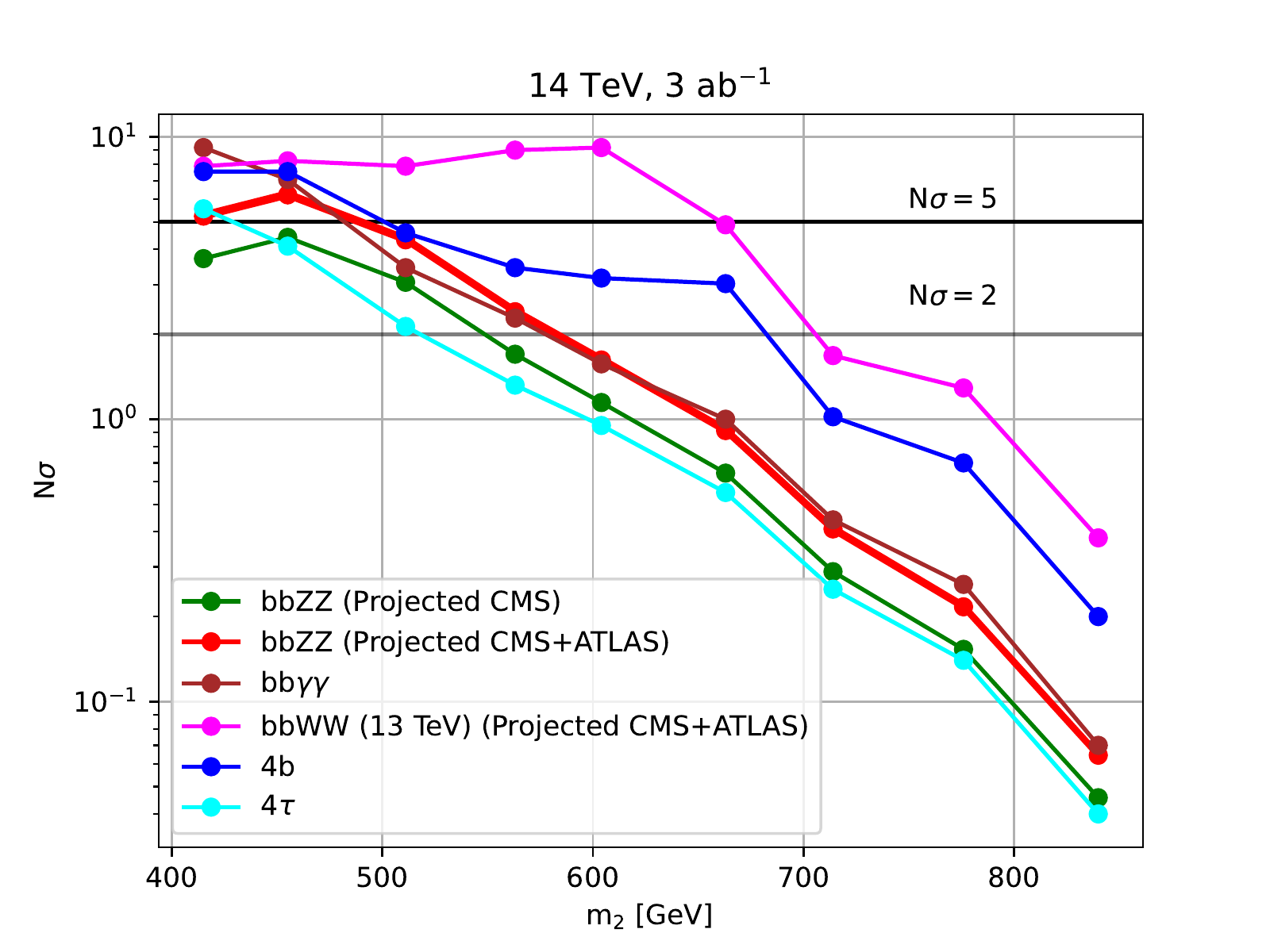}
    \caption{Comparison of significance, $\rm{N}_\sigma$, as a function of resonance mass corresponding to different benchmark points as in Table \ref{tab:bmpoint} for different channels for 14 TeV HL-LHC with an integrated luminosity of $3 ~\rm{ab}^{-1}$ is shown. The values of our channel $bbZZ$ are shown by green and red line for projection of CMS experiment and for projection of combination of CMS and ATLAS experiments, respectively. The values corresponding to $bb\gamma \gamma$, $bbWW$, $4\tau$ and $4b$ channels are shown by brown, magenta, blue and cyan lines, respectively and these values are taken from the Refs. \cite{Kotwal:2016tex, Huang:2017jws, Li:2019tfd}.}
     \label{fig:comparesig}
\end{figure*}

To evaluate the sensitivity, the $\rm CLs$ value is estimated 
from the distributions of final parametrized DNN score with the profile likelihood method considering the asymptotic formula explained in Refs. \cite{Cowan:2013pha, Cowan:2010js}. $\rm{N}_{\sigma}$ Gaussian significance is calculated by converting $\rm{CLs}$ values. The uncertainty band influenced by the systematic uncertainties considered is made by altering the event yields for the uncertainties of $1\sigma$, computing the $\rm{CLs}$ and converting it to significance as mentioned above, for instance, for BM4 of mass $\rm{m}_{2} = 455$ GeV, the uncertainty on significance is $(+2.2 \%, - 1.3 \%)$.
\paragraph*{}$\rm{N}_\sigma$ is shown as a function of resonance mass in Fig. \ref{fig:comparesig}. 
The significance corresponds to a combination of $\mu \mu$ and $ee$ channel, taking the efficiencies of signal and background selection of $ee$ channel to be equal to those of the $\mu \mu$ channel estimated in the analysis above.
If we search with only data from CMS experiment with $3~ \rm{ab^{-1}}$ of integrated luminosity at the 14 TeV HL-LHC, discovery significance can not be observed over the range of resonant masses, though if a signal is not observed in the future HL-LHC experiments, the maximum signal rate BM points up to $\rm{m}_{2} = 550$ GeV at $95\%$ C.L. can be excluded. If we presume to combine the data from CMS and ATLAS experiments eventually, by doubling the number of events obtained in this analysis, it is possible to discover the BM points having maximum signal rate up to $m_2 = 490$ GeV with $\rm{N}_\sigma >$ 5. Also, if we do not observe a signal in the future HL-LHC experiments, the maximum signal rate BM points up to $\rm{m}_{2} = 580$ GeV can be excluded at $95\%$ C.L.
\paragraph*{}We compare the significance with those of $bb\gamma \gamma$, $4\tau$ and $4b$ channels for the 14 TeV HL-LHC and in case of $bbWW$ channel for the 13 TeV LHC and it is shown in Fig. \ref{fig:comparesig}. The benchmark points are only compared from BM3 to BM11 because there is difference in BM1 and BM2 points with respect to the Ref. \cite{Kotwal:2016tex}. In case of heavy Higgs mass,
$\rm{m}_2$ less than 490 GeV, for a search of resonant di-Higgs production, the maximum sensitivity is found in the $4b$ final state, if we ignore the 13 TeV CMS-ATLAS combination of bbWW channel. For $\rm{m}_2$ above 490 GeV, the combination of CMS and ATLAS experiments gives the estimates of the significance for the $bbZZ$ final state quite comparable with $bb\gamma \gamma$, even better at some BM points. In addition, the $bbZZ$ channel with a combination of CMS and ATLAS data can be a potential probe with the $bb\gamma \gamma$ and $4b$ channel, which can be used as a complementary check if we observe a signal in the $bb\gamma \gamma$ and $4b$ channels. Also, it has a higher sensitivity than that of $4\tau$ channel. The $bbWW$ analysis does not consider the Drell-Yan in association with jets background where significant contribution is expected from.
 \section{Discussion \& Summary}
 \label{summary}
 In this paper, we observed the prospect of HL-LHC in discovering a resonant production of the heavy singlet-like scalar by gluon fusion process in the xSM which eventually decays into SM-like Higgs pair in $bbZZ$ final state, where one $Z$ boson decays into two electrons or muons and another decays to two neutrinos. This final state is important, where it balances between its large branching ratio in the $bb$ channel having a huge amount of QCD background and the clean leptonic decay of $Z$ along with significant branching ratio of $Z$ in decaying to neutrinos. To validate the analysis, our simulation was compared against the CMS 13 TeV analysis.  For a heavy singlet-like scalar mass from 300 to 850 GeV, we selected 11 benchmark points that satisfy a SFOEWPT along with the theoretical and phenomenological constraints and calculated the significance at the 14 TeV HL-LHC for an integrated luminosity of $3 ~\rm{ab}^{-1}$. For that purpose, a parameterized DNN was used for final discrimination, which is an efficient method that helps for the classification of the signal and background processes by utilizing only a single network for all the benchmark points. For a phenomenological analysis, we used parameterised DNN for the first time for final event discrimination and we encourage to use this simple and effective model for other phenomenological analyses involving a number of parameters corresponding to different signal hypotheses. 
 Our results are also compared with the previous analyses in $bb\gamma \gamma$ , $bbWW$, $4b$ and  $4\tau$ channels and it was concluded that inclusion of $bbZZ(\ell \ell\nu \nu)$ 
could be a potential search channel for a combination of CMS and ATLAS experiments along with the other channels previously explored. 
 Also, we can exclude regions of parameter space if HL-LHC discovery is not achieved. However, in case of the benchmark points corresponding to minimum signal rate,
exclusion of a signal will never be possible.
 Therefore, if the chance of SFOEWPT generation in the xSM needs to be fully excluded, it may require a 100 TeV pp collider in future.
 
 \section{Acknowledgments}
 The authors would like to thank Anish Ghoshal, Manimala Mitra, Haolin Li, Debabrata Bhowmik, Gourab Saha and Arnab Roy for useful discussions and Subir Sarkar for computing infrastructure support. PP is supported by Senior Research Fellowship from University Grant Commission (UGC), India. SS is supported by FAPESP grant 2021/09547-9. 

\appendix
\section{Descriptions of kinematic variables}
\label{apndx1}
\begin{figure*}[h!]
  \centering
 \includegraphics[width=0.99\textwidth]{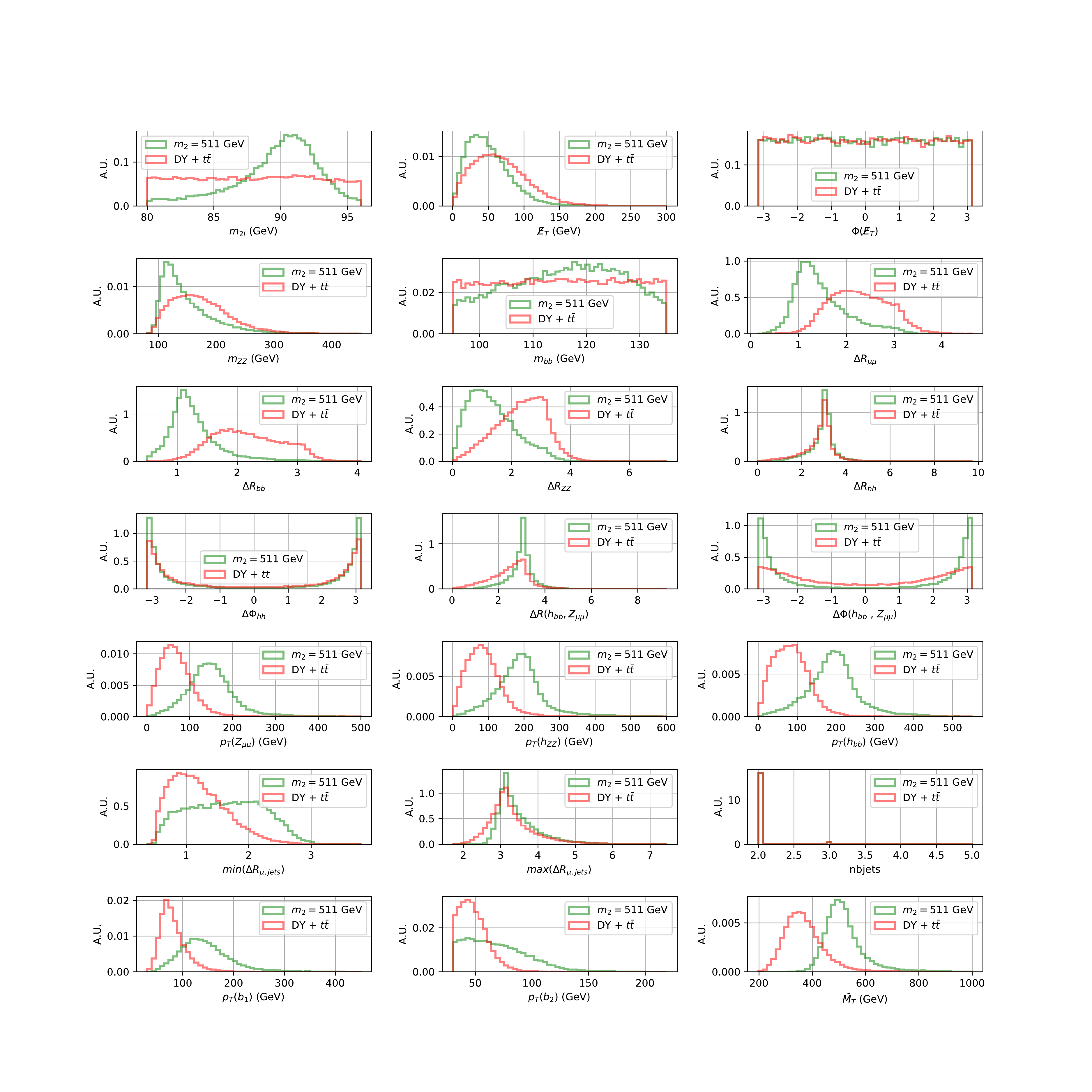}
 \caption{Distributions of the input variables to parameterized DNN normalized to unit area are shown. The signal corresponds to BM5 with $m_2$ of $511$ GeV. The signal and background distributions are represented by the green and red lines, respectively.}
  \label{fig:varsnorm}
\end{figure*}

Descriptions of the kinematic variables which are given as input to the parameterized DNN classifier are given in Table \ref{table:vardnn} below. We plot the distributions of the input variables in Fig. \ref{fig:varsnorm} for signal and combined background of $t\bar{t}$ and DY used for 14 TeV prediction. The signal is taken corresponding to BM5 given in Table \ref{tab:bmpoint}. All the plots are normalized to unit area. 
\begin{table}[hbt!]
\begin{center}
\begin{tabular}{ |c|c| }
 \hline
 Variables & Descriptions \\
\hline
    $m_{2\ell}$  & Invariant mass of the pair of leptons \\
 \hline
    $m_{b b}$ & Invariant mass of the pair of b-jets \\
  \hline
    $m_{Z Z}$ & Invariant mass of the pair of $Z$ boson candidates \\
  \hline 
    $\slashed{E}_T$ & Missing transverse momentum\\
  \hline
    $\Phi(\slashed{E}_{T})$ & $\Phi$ co-ordinates of the missing transverse momentum four vector\\
  \hline 
    $\Delta R_{b b}$ & $\Delta R$ between two leading b-jets\\
  \hline  
    $\Delta R_{\mu \mu}$ & $\Delta R$ between two leading muons\\
  \hline  
    $\Delta R_{Z Z}$ & $\Delta R$ between the pair of $Z$ bosons\\
  \hline
    $\Delta R_{h h}$ & $\Delta R$ between the pair of Higgs bosons\\
  \hline
    $\Delta R(h_{b b}, Z_{\mu \mu})$ & $\Delta R$ between the Higgs reconstructed from pair of b-jets and $Z$ boson reconstructed from pair of leptons\\
  \hline
    $\Delta \Phi_{h h}$ & $\Delta \Phi$ between the pair of Higgs bosons\\
  \hline
    $\Delta \Phi(h_{b b}$ , $Z_{\mu \mu})$ & $\Delta \Phi$ between the Higgs reconstructed from pair of b-jets and $Z$ boson reconstructed from pair of leptons\\
  \hline
    $p_T (Z_{\mu \mu})$ & Transverse momentum of the $Z$ bosons reconstructed from pair of muons\\
  \hline
    $p_T (h_{Z Z})$ & Transverse momentum of the Higgs bosons reconstructed from pair of $Z$ bosons\\
  \hline
    $p_T (h_{b b})$ & Transverse momentum of the Higgs bosons reconstructed from pair of b-jets\\
  \hline
    $ min(\Delta R_{\mu , jets})$ & Minimum of $\Delta R$ between the muons and jets in the event \\
  \hline 
     $ max(\Delta R_{\mu , jets})$ & Maximum of $\Delta R$ between the muons and jets in the event \\
  \hline 
     nbjets & Number of b-jets in the event\\
  \hline 
     $p_T ( b_1)$ & Transverse momentum of the leading b-jet\\
   \hline
     $p_T ( b_2)$ & Transverse momentum of the sub-leading b-jet\\
   \hline
     $\Tilde{M}_T$ & Pseudo transverse mass of the di-Higgs candidate\\
   \hline       
\end{tabular}
\caption{Descriptions of the kinematic variables used as input variables to the parameterized DNN}
\label{table:vardnn}
\end{center}
\end{table}

\clearpage
\section{BDT and Neural network architectures}
\label{apndx2}
\subsubsection{BDT training for 13 TeV analysis}
TMVA \cite{TMVA:2007ngy} is used for BDT \cite{10.1214/aos/1013203451, freund1997decision} training. 
\begin{table}[h]
\centering
\begin{tabular}{|l|l|}
\hline
\textbf{Parameter} & \textbf{Value} \\ \hline
NTrees & 850 \\ 
MinNodeSize & 2.5\% \\ 
MaxDepth & 3 \\ 
BoostType & AdaBoost \\ 
AdaBoostBeta & 0.5 \\ 
BaggedSampleFraction & 0.5 \\ 
SeparationType & GiniIndex \\ 
nCuts & 20 \\ \hline
\end{tabular}
\caption{Hyperparameters used in BDT training}
\label{tab:ada_boost_params}
\end{table}

\subsubsection{Parameterized DNN training for 14 TeV analysis}

Keras \cite{chollet2015keras} is used for parameterized DNN training. 

\begin{table}[h]
\centering
\begin{tabular}{|l|l|}
\hline
\textbf{Parameter} & \textbf{Value} \\ \hline
Feature scaling & MinMaxScaler \cite{scikit-learn}\\ 
Activation function & ReLU ]\cite{NairHinton2010}  in all layers, except Softmax \cite{Bridle1990} in final output layer \\ 
Optimizer & Adam \cite{KingmaBa2014}  \\ 
Loss function & Categorical crossentropy \cite{RubinsteinKroese2004} \\ 
Metrics & Accuracy \cite{SokolovaLapalme2009}\\ 
Epochs & 100 \\ \hline
\end{tabular}
\caption{Hyperparameters used in Parameterized DNN training}
\label{tab:nn_config}
\end{table}


\begin{table}[h]
\centering
\begin{tabular}{|l|l|l|}
\hline
\textbf{Model}         & \multicolumn{2}{l|}{\textbf{Sequential}}    \\ \hline
\textbf{Layer (type)}  & \textbf{Output Shape} & \textbf{Param \#} \\ \hline
dense\_1 (Dense)                 & (None, 70)             & 1,820              \\ \hline
dense\_2  (Dense)               & (None, 50)             & 3,550              \\ \hline
dense\_3  (Dense)               & (None, 30)             & 1,530              \\ \hline
dense\_4  (Dense)               & (None, 20)             & 620                \\ \hline
dense\_5  (Dense)               & (None, 10)             & 210                \\ \hline
dense\_6  (Dense)              & (None, 2)              & 22                 \\ \hline
\textbf{Total params}   & \multicolumn{2}{l|}{7,752 (30.28 KB)}       \\ \hline
\textbf{Trainable params} & \multicolumn{2}{l|}{7,752 (30.28 KB)}    \\ \hline
\textbf{Non-trainable params} & \multicolumn{2}{l|}{0 (0.00 B)}       \\ \hline
\end{tabular}
\caption{Model Architecture Summary}
\label{tab:model_architecture}
\end{table}

\clearpage
\bibliographystyle{apsrev}
\bibliography{bibliography_new}

\begin{thebibliography}{84}
\expandafter\ifx\csname natexlab\endcsname\relax\def\natexlab#1{#1}\fi
\expandafter\ifx\csname bibnamefont\endcsname\relax
  \def\bibnamefont#1{#1}\fi
\expandafter\ifx\csname bibfnamefont\endcsname\relax
  \def\bibfnamefont#1{#1}\fi
\expandafter\ifx\csname citenamefont\endcsname\relax
  \def\citenamefont#1{#1}\fi
\expandafter\ifx\csname url\endcsname\relax
  \def\url#1{\texttt{#1}}\fi
\expandafter\ifx\csname urlprefix\endcsname\relax\def\urlprefix{URL }\fi
\providecommand{\bibinfo}[2]{#2}
\providecommand{\eprint}[2][]{\url{#2}}

\bibitem[{\citenamefont{Chatrchyan et~al.}(2012)}]{CMS:2012qbp}
\bibinfo{author}{\bibfnamefont{S.}~\bibnamefont{Chatrchyan}}
  \bibnamefont{et~al.} (\bibinfo{collaboration}{CMS}), \bibinfo{journal}{Phys.
  Lett. B} \textbf{\bibinfo{volume}{716}}, \bibinfo{pages}{30}
  (\bibinfo{year}{2012}), \eprint{1207.7235}.

\bibitem[{\citenamefont{Aad et~al.}(2012)}]{ATLAS:2012yve}
\bibinfo{author}{\bibfnamefont{G.}~\bibnamefont{Aad}} \bibnamefont{et~al.}
  (\bibinfo{collaboration}{ATLAS}), \bibinfo{journal}{Phys. Lett. B}
  \textbf{\bibinfo{volume}{716}}, \bibinfo{pages}{1} (\bibinfo{year}{2012}),
  \eprint{1207.7214}.

\bibitem[{\citenamefont{Ade et~al.}(2014)}]{Planck:2013pxb}
\bibinfo{author}{\bibfnamefont{P.~A.~R.} \bibnamefont{Ade}}
  \bibnamefont{et~al.} (\bibinfo{collaboration}{Planck}),
  \bibinfo{journal}{Astron. Astrophys.} \textbf{\bibinfo{volume}{571}},
  \bibinfo{pages}{A16} (\bibinfo{year}{2014}), \eprint{1303.5076}.

\bibitem[{\citenamefont{Akhmedov et~al.}(2000)\citenamefont{Akhmedov, Branco,
  and Rebelo}}]{AKHMEDOV2000215}
\bibinfo{author}{\bibfnamefont{E.}~\bibnamefont{Akhmedov}},
  \bibinfo{author}{\bibfnamefont{G.}~\bibnamefont{Branco}}, \bibnamefont{and}
  \bibinfo{author}{\bibfnamefont{M.}~\bibnamefont{Rebelo}},
  \bibinfo{journal}{Physics Letters B} \textbf{\bibinfo{volume}{478}},
  \bibinfo{pages}{215} (\bibinfo{year}{2000}), ISSN \bibinfo{issn}{0370-2693},
  \urlprefix\url{https://www.sciencedirect.com/science/article/pii/S0370269300002823}.

\bibitem[{\citenamefont{{Zwicky}}(1933)}]{1933AcHPh...6..110Z}
\bibinfo{author}{\bibfnamefont{F.}~\bibnamefont{{Zwicky}}},
  \bibinfo{journal}{Helvetica Physica Acta} \textbf{\bibinfo{volume}{6}},
  \bibinfo{pages}{110} (\bibinfo{year}{1933}).

\bibitem[{\citenamefont{{Zwicky}}(1937)}]{1937ApJ....86..217Z}
\bibinfo{author}{\bibfnamefont{F.}~\bibnamefont{{Zwicky}}},
  \bibinfo{journal}{\apj} \textbf{\bibinfo{volume}{86}}, \bibinfo{pages}{217}
  (\bibinfo{year}{1937}).

\bibitem[{\citenamefont{Sakharov}(1967)}]{Sakharov:1967dj}
\bibinfo{author}{\bibfnamefont{A.~D.} \bibnamefont{Sakharov}},
  \bibinfo{journal}{Pisma Zh. Eksp. Teor. Fiz.} \textbf{\bibinfo{volume}{5}},
  \bibinfo{pages}{32} (\bibinfo{year}{1967}).

\bibitem[{\citenamefont{Aoki et~al.}(1999)\citenamefont{Aoki, Csikor, Fodor,
  and Ukawa}}]{Aoki:1999fi}
\bibinfo{author}{\bibfnamefont{Y.}~\bibnamefont{Aoki}},
  \bibinfo{author}{\bibfnamefont{F.}~\bibnamefont{Csikor}},
  \bibinfo{author}{\bibfnamefont{Z.}~\bibnamefont{Fodor}}, \bibnamefont{and}
  \bibinfo{author}{\bibfnamefont{A.}~\bibnamefont{Ukawa}},
  \bibinfo{journal}{Phys. Rev. D} \textbf{\bibinfo{volume}{60}},
  \bibinfo{pages}{013001} (\bibinfo{year}{1999}), \eprint{hep-lat/9901021}.

\bibitem[{\citenamefont{Csikor et~al.}(1999)\citenamefont{Csikor, Fodor, and
  Heitger}}]{Csikor:1998eu}
\bibinfo{author}{\bibfnamefont{F.}~\bibnamefont{Csikor}},
  \bibinfo{author}{\bibfnamefont{Z.}~\bibnamefont{Fodor}}, \bibnamefont{and}
  \bibinfo{author}{\bibfnamefont{J.}~\bibnamefont{Heitger}},
  \bibinfo{journal}{Phys. Rev. Lett.} \textbf{\bibinfo{volume}{82}},
  \bibinfo{pages}{21} (\bibinfo{year}{1999}), \eprint{hep-ph/9809291}.

\bibitem[{\citenamefont{Laine and Rummukainen}(1999)}]{Laine:1998jb}
\bibinfo{author}{\bibfnamefont{M.}~\bibnamefont{Laine}} \bibnamefont{and}
  \bibinfo{author}{\bibfnamefont{K.}~\bibnamefont{Rummukainen}},
  \bibinfo{journal}{Nucl. Phys. B Proc. Suppl.} \textbf{\bibinfo{volume}{73}},
  \bibinfo{pages}{180} (\bibinfo{year}{1999}), \eprint{hep-lat/9809045}.

\bibitem[{\citenamefont{Gurtler et~al.}(1997)\citenamefont{Gurtler, Ilgenfritz,
  and Schiller}}]{Gurtler:1997hr}
\bibinfo{author}{\bibfnamefont{M.}~\bibnamefont{Gurtler}},
  \bibinfo{author}{\bibfnamefont{E.-M.} \bibnamefont{Ilgenfritz}},
  \bibnamefont{and} \bibinfo{author}{\bibfnamefont{A.}~\bibnamefont{Schiller}},
  \bibinfo{journal}{Phys. Rev. D} \textbf{\bibinfo{volume}{56}},
  \bibinfo{pages}{3888} (\bibinfo{year}{1997}), \eprint{hep-lat/9704013}.

\bibitem[{\citenamefont{Profumo et~al.}(2007)\citenamefont{Profumo,
  Ramsey-Musolf, and Shaughnessy}}]{Profumo:2007wc}
\bibinfo{author}{\bibfnamefont{S.}~\bibnamefont{Profumo}},
  \bibinfo{author}{\bibfnamefont{M.~J.} \bibnamefont{Ramsey-Musolf}},
  \bibnamefont{and}
  \bibinfo{author}{\bibfnamefont{G.}~\bibnamefont{Shaughnessy}},
  \bibinfo{journal}{JHEP} \textbf{\bibinfo{volume}{08}}, \bibinfo{pages}{010}
  (\bibinfo{year}{2007}), \eprint{0705.2425}.

\bibitem[{\citenamefont{Espinosa et~al.}(2012)\citenamefont{Espinosa,
  Konstandin, and Riva}}]{Espinosa:2011ax}
\bibinfo{author}{\bibfnamefont{J.~R.} \bibnamefont{Espinosa}},
  \bibinfo{author}{\bibfnamefont{T.}~\bibnamefont{Konstandin}},
  \bibnamefont{and} \bibinfo{author}{\bibfnamefont{F.}~\bibnamefont{Riva}},
  \bibinfo{journal}{Nucl. Phys. B} \textbf{\bibinfo{volume}{854}},
  \bibinfo{pages}{592} (\bibinfo{year}{2012}), \eprint{1107.5441}.

\bibitem[{\citenamefont{Ghorbani and Ghorbani}(2020)}]{Ghorbani:2018yfr}
\bibinfo{author}{\bibfnamefont{K.}~\bibnamefont{Ghorbani}} \bibnamefont{and}
  \bibinfo{author}{\bibfnamefont{P.~H.} \bibnamefont{Ghorbani}},
  \bibinfo{journal}{J. Phys. G} \textbf{\bibinfo{volume}{47}},
  \bibinfo{pages}{015201} (\bibinfo{year}{2020}), \eprint{1804.05798}.

\bibitem[{\citenamefont{Tumasyan et~al.}(2022)}]{CMS:2022dwd}
\bibinfo{author}{\bibfnamefont{A.}~\bibnamefont{Tumasyan}} \bibnamefont{et~al.}
  (\bibinfo{collaboration}{CMS}), \bibinfo{journal}{Nature}
  \textbf{\bibinfo{volume}{607}}, \bibinfo{pages}{60} (\bibinfo{year}{2022}),
  \eprint{2207.00043}.

\bibitem[{\citenamefont{Aad et~al.}(2022)}]{ATLAS:2022vkf}
\bibinfo{author}{\bibfnamefont{G.}~\bibnamefont{Aad}} \bibnamefont{et~al.}
  (\bibinfo{collaboration}{ATLAS}), \bibinfo{journal}{Nature}
  \textbf{\bibinfo{volume}{607}}, \bibinfo{pages}{52} (\bibinfo{year}{2022}),
  \bibinfo{note}{[Erratum: Nature 612, E24 (2022)]}, \eprint{2207.00092}.

\bibitem[{\citenamefont{Aaboud et~al.}(2019{\natexlab{a}})}]{ATLAS:2018rnh}
\bibinfo{author}{\bibfnamefont{M.}~\bibnamefont{Aaboud}} \bibnamefont{et~al.}
  (\bibinfo{collaboration}{ATLAS}), \bibinfo{journal}{JHEP}
  \textbf{\bibinfo{volume}{01}}, \bibinfo{pages}{030}
  (\bibinfo{year}{2019}{\natexlab{a}}), \eprint{1804.06174}.

\bibitem[{\citenamefont{Sirunyan et~al.}(2018{\natexlab{a}})}]{CMS:2018qmt}
\bibinfo{author}{\bibfnamefont{A.~M.} \bibnamefont{Sirunyan}}
  \bibnamefont{et~al.} (\bibinfo{collaboration}{CMS}), \bibinfo{journal}{JHEP}
  \textbf{\bibinfo{volume}{08}}, \bibinfo{pages}{152}
  (\bibinfo{year}{2018}{\natexlab{a}}), \eprint{1806.03548}.

\bibitem[{\citenamefont{Aaboud et~al.}(2019{\natexlab{b}})}]{ATLAS:2018fpd}
\bibinfo{author}{\bibfnamefont{M.}~\bibnamefont{Aaboud}} \bibnamefont{et~al.}
  (\bibinfo{collaboration}{ATLAS}), \bibinfo{journal}{JHEP}
  \textbf{\bibinfo{volume}{04}}, \bibinfo{pages}{092}
  (\bibinfo{year}{2019}{\natexlab{b}}), \eprint{1811.04671}.

\bibitem[{\citenamefont{Sirunyan et~al.}(2018{\natexlab{b}})}]{CMS:2017rpp}
\bibinfo{author}{\bibfnamefont{A.~M.} \bibnamefont{Sirunyan}}
  \bibnamefont{et~al.} (\bibinfo{collaboration}{CMS}), \bibinfo{journal}{JHEP}
  \textbf{\bibinfo{volume}{01}}, \bibinfo{pages}{054}
  (\bibinfo{year}{2018}{\natexlab{b}}), \eprint{1708.04188}.

\bibitem[{\citenamefont{Sirunyan et~al.}(2020)\citenamefont{Sirunyan, Tumasyan,
  Adam, Ambrogi, Bergauer, Dragicevic, Er\"o, Escalante Del~Valle, Fr\"uhwirth,
  Jeitler et~al.}}]{PhysRevD.102.032003}
\bibinfo{author}{\bibfnamefont{A.~M.} \bibnamefont{Sirunyan}},
  \bibinfo{author}{\bibfnamefont{A.}~\bibnamefont{Tumasyan}},
  \bibinfo{author}{\bibfnamefont{W.}~\bibnamefont{Adam}},
  \bibinfo{author}{\bibfnamefont{F.}~\bibnamefont{Ambrogi}},
  \bibinfo{author}{\bibfnamefont{T.}~\bibnamefont{Bergauer}},
  \bibinfo{author}{\bibfnamefont{M.}~\bibnamefont{Dragicevic}},
  \bibinfo{author}{\bibfnamefont{J.}~\bibnamefont{Er\"o}},
  \bibinfo{author}{\bibfnamefont{A.}~\bibnamefont{Escalante Del~Valle}},
  \bibinfo{author}{\bibfnamefont{R.}~\bibnamefont{Fr\"uhwirth}},
  \bibinfo{author}{\bibfnamefont{M.}~\bibnamefont{Jeitler}},
  \bibnamefont{et~al.} (\bibinfo{collaboration}{CMS Collaboration}),
  \bibinfo{journal}{Phys. Rev. D} \textbf{\bibinfo{volume}{102}},
  \bibinfo{pages}{032003} (\bibinfo{year}{2020}),
  \urlprefix\url{https://link.aps.org/doi/10.1103/PhysRevD.102.032003}.

\bibitem[{\citenamefont{Aaboud et~al.}(2018{\natexlab{a}})}]{ATLAS:2018uni}
\bibinfo{author}{\bibfnamefont{M.}~\bibnamefont{Aaboud}} \bibnamefont{et~al.}
  (\bibinfo{collaboration}{ATLAS}), \bibinfo{journal}{Phys. Rev. Lett.}
  \textbf{\bibinfo{volume}{121}}, \bibinfo{pages}{191801}
  (\bibinfo{year}{2018}{\natexlab{a}}), \bibinfo{note}{[Erratum: Phys.Rev.Lett.
  122, 089901 (2019)]}, \eprint{1808.00336}.

\bibitem[{\citenamefont{Sirunyan et~al.}(2018{\natexlab{c}})}]{CMS:2017hea}
\bibinfo{author}{\bibfnamefont{A.~M.} \bibnamefont{Sirunyan}}
  \bibnamefont{et~al.} (\bibinfo{collaboration}{CMS}), \bibinfo{journal}{Phys.
  Lett. B} \textbf{\bibinfo{volume}{778}}, \bibinfo{pages}{101}
  (\bibinfo{year}{2018}{\natexlab{c}}), \eprint{1707.02909}.

\bibitem[{\citenamefont{Aaboud et~al.}(2018{\natexlab{b}})}]{ATLAS:2018dpp}
\bibinfo{author}{\bibfnamefont{M.}~\bibnamefont{Aaboud}} \bibnamefont{et~al.}
  (\bibinfo{collaboration}{ATLAS}), \bibinfo{journal}{JHEP}
  \textbf{\bibinfo{volume}{11}}, \bibinfo{pages}{040}
  (\bibinfo{year}{2018}{\natexlab{b}}), \eprint{1807.04873}.

\bibitem[{\citenamefont{Sirunyan et~al.}(2019)}]{CMS:2018tla}
\bibinfo{author}{\bibfnamefont{A.~M.} \bibnamefont{Sirunyan}}
  \bibnamefont{et~al.} (\bibinfo{collaboration}{CMS}), \bibinfo{journal}{Phys.
  Lett. B} \textbf{\bibinfo{volume}{788}}, \bibinfo{pages}{7}
  (\bibinfo{year}{2019}), \eprint{1806.00408}.

\bibitem[{\citenamefont{Aaboud et~al.}(2019{\natexlab{c}})}]{ATLAS:2018ili}
\bibinfo{author}{\bibfnamefont{M.}~\bibnamefont{Aaboud}} \bibnamefont{et~al.}
  (\bibinfo{collaboration}{ATLAS}), \bibinfo{journal}{JHEP}
  \textbf{\bibinfo{volume}{05}}, \bibinfo{pages}{124}
  (\bibinfo{year}{2019}{\natexlab{c}}), \eprint{1811.11028}.

\bibitem[{\citenamefont{Aaboud et~al.}(2018{\natexlab{c}})}]{ATLAS:2018hqk}
\bibinfo{author}{\bibfnamefont{M.}~\bibnamefont{Aaboud}} \bibnamefont{et~al.}
  (\bibinfo{collaboration}{ATLAS}), \bibinfo{journal}{Eur. Phys. J. C}
  \textbf{\bibinfo{volume}{78}}, \bibinfo{pages}{1007}
  (\bibinfo{year}{2018}{\natexlab{c}}), \eprint{1807.08567}.

\bibitem[{\citenamefont{Kotwal et~al.}(2016)\citenamefont{Kotwal,
  Ramsey-Musolf, No, and Winslow}}]{Kotwal:2016tex}
\bibinfo{author}{\bibfnamefont{A.~V.} \bibnamefont{Kotwal}},
  \bibinfo{author}{\bibfnamefont{M.~J.} \bibnamefont{Ramsey-Musolf}},
  \bibinfo{author}{\bibfnamefont{J.~M.} \bibnamefont{No}}, \bibnamefont{and}
  \bibinfo{author}{\bibfnamefont{P.}~\bibnamefont{Winslow}},
  \bibinfo{journal}{Phys. Rev. D} \textbf{\bibinfo{volume}{94}},
  \bibinfo{pages}{035022} (\bibinfo{year}{2016}), \eprint{1605.06123}.

\bibitem[{\citenamefont{Li et~al.}(2019)\citenamefont{Li, Ramsey-Musolf, and
  Willocq}}]{Li:2019tfd}
\bibinfo{author}{\bibfnamefont{H.-L.} \bibnamefont{Li}},
  \bibinfo{author}{\bibfnamefont{M.}~\bibnamefont{Ramsey-Musolf}},
  \bibnamefont{and} \bibinfo{author}{\bibfnamefont{S.}~\bibnamefont{Willocq}},
  \bibinfo{journal}{Phys. Rev. D} \textbf{\bibinfo{volume}{100}},
  \bibinfo{pages}{075035} (\bibinfo{year}{2019}), \eprint{1906.05289}.

\bibitem[{\citenamefont{Huang et~al.}(2017)\citenamefont{Huang, No, Perni\'e,
  Ramsey-Musolf, Safonov, Spannowsky, and Winslow}}]{Huang:2017jws}
\bibinfo{author}{\bibfnamefont{T.}~\bibnamefont{Huang}},
  \bibinfo{author}{\bibfnamefont{J.~M.} \bibnamefont{No}},
  \bibinfo{author}{\bibfnamefont{L.}~\bibnamefont{Perni\'e}},
  \bibinfo{author}{\bibfnamefont{M.}~\bibnamefont{Ramsey-Musolf}},
  \bibinfo{author}{\bibfnamefont{A.}~\bibnamefont{Safonov}},
  \bibinfo{author}{\bibfnamefont{M.}~\bibnamefont{Spannowsky}},
  \bibnamefont{and} \bibinfo{author}{\bibfnamefont{P.}~\bibnamefont{Winslow}},
  \bibinfo{journal}{Phys. Rev. D} \textbf{\bibinfo{volume}{96}},
  \bibinfo{pages}{035007} (\bibinfo{year}{2017}), \eprint{1701.04442}.

\bibitem[{\citenamefont{L\"u et~al.}(2016)\citenamefont{L\"u, Du, Fang, He, and
  Zhang}}]{Lu:2015qqa}
\bibinfo{author}{\bibfnamefont{L.-C.} \bibnamefont{L\"u}},
  \bibinfo{author}{\bibfnamefont{C.}~\bibnamefont{Du}},
  \bibinfo{author}{\bibfnamefont{Y.}~\bibnamefont{Fang}},
  \bibinfo{author}{\bibfnamefont{H.-J.} \bibnamefont{He}}, \bibnamefont{and}
  \bibinfo{author}{\bibfnamefont{H.}~\bibnamefont{Zhang}},
  \bibinfo{journal}{Phys. Lett. B} \textbf{\bibinfo{volume}{755}},
  \bibinfo{pages}{509} (\bibinfo{year}{2016}), \eprint{1507.02644}.

\bibitem[{\citenamefont{Ren et~al.}(2018)\citenamefont{Ren, Xiao, Zhou, Fang,
  He, and Yao}}]{Ren:2017jbg}
\bibinfo{author}{\bibfnamefont{J.}~\bibnamefont{Ren}},
  \bibinfo{author}{\bibfnamefont{R.-Q.} \bibnamefont{Xiao}},
  \bibinfo{author}{\bibfnamefont{M.}~\bibnamefont{Zhou}},
  \bibinfo{author}{\bibfnamefont{Y.}~\bibnamefont{Fang}},
  \bibinfo{author}{\bibfnamefont{H.-J.} \bibnamefont{He}}, \bibnamefont{and}
  \bibinfo{author}{\bibfnamefont{W.}~\bibnamefont{Yao}},
  \bibinfo{journal}{JHEP} \textbf{\bibinfo{volume}{06}}, \bibinfo{pages}{090}
  (\bibinfo{year}{2018}), \eprint{1706.05980}.

\bibitem[{CMS(2018{\natexlab{a}})}]{CMS-PAS-HIG-17-032}
\bibinfo{type}{Tech. Rep.}, \bibinfo{institution}{(CMS)},
  \bibinfo{address}{Geneva} (\bibinfo{year}{2018}{\natexlab{a}}),
  \urlprefix\url{http://cds.cern.ch/record/2648796}.

\bibitem[{\citenamefont{Alves et~al.}(2018)\citenamefont{Alves, Ghosh, Guo, and
  Sinha}}]{Alves:2018oct}
\bibinfo{author}{\bibfnamefont{A.}~\bibnamefont{Alves}},
  \bibinfo{author}{\bibfnamefont{T.}~\bibnamefont{Ghosh}},
  \bibinfo{author}{\bibfnamefont{H.-K.} \bibnamefont{Guo}}, \bibnamefont{and}
  \bibinfo{author}{\bibfnamefont{K.}~\bibnamefont{Sinha}},
  \bibinfo{journal}{JHEP} \textbf{\bibinfo{volume}{12}}, \bibinfo{pages}{070}
  (\bibinfo{year}{2018}), \eprint{1808.08974}.

\bibitem[{\citenamefont{Alves et~al.}(2019)\citenamefont{Alves, Ghosh, and
  Queiroz}}]{Alves:2019emf}
\bibinfo{author}{\bibfnamefont{A.}~\bibnamefont{Alves}},
  \bibinfo{author}{\bibfnamefont{T.}~\bibnamefont{Ghosh}}, \bibnamefont{and}
  \bibinfo{author}{\bibfnamefont{F.~S.} \bibnamefont{Queiroz}},
  \bibinfo{journal}{Phys. Rev. D} \textbf{\bibinfo{volume}{100}},
  \bibinfo{pages}{036012} (\bibinfo{year}{2019}), \eprint{1905.03271}.

\bibitem[{\citenamefont{Baldi et~al.}(2016)\citenamefont{Baldi, Cranmer,
  Faucett, Sadowski, and Whiteson}}]{Baldi:2016fzo}
\bibinfo{author}{\bibfnamefont{P.}~\bibnamefont{Baldi}},
  \bibinfo{author}{\bibfnamefont{K.}~\bibnamefont{Cranmer}},
  \bibinfo{author}{\bibfnamefont{T.}~\bibnamefont{Faucett}},
  \bibinfo{author}{\bibfnamefont{P.}~\bibnamefont{Sadowski}}, \bibnamefont{and}
  \bibinfo{author}{\bibfnamefont{D.}~\bibnamefont{Whiteson}},
  \bibinfo{journal}{Eur. Phys. J. C} \textbf{\bibinfo{volume}{76}},
  \bibinfo{pages}{235} (\bibinfo{year}{2016}), \eprint{1601.07913}.

\bibitem[{\citenamefont{Anzalone et~al.}(2022)\citenamefont{Anzalone,
  Diotalevi, and Bonacorsi}}]{Anzalone_2022}
\bibinfo{author}{\bibfnamefont{L.}~\bibnamefont{Anzalone}},
  \bibinfo{author}{\bibfnamefont{T.}~\bibnamefont{Diotalevi}},
  \bibnamefont{and}
  \bibinfo{author}{\bibfnamefont{D.}~\bibnamefont{Bonacorsi}},
  \bibinfo{journal}{Machine Learning: Science and Technology}
  \textbf{\bibinfo{volume}{3}}, \bibinfo{pages}{035017} (\bibinfo{year}{2022}),
  \urlprefix\url{https://dx.doi.org/10.1088/2632-2153/ac917c}.

\bibitem[{\citenamefont{Alves et~al.}(2020)\citenamefont{Alves, Gon\c{c}alves,
  Ghosh, Guo, and Sinha}}]{Alves:2019igs}
\bibinfo{author}{\bibfnamefont{A.}~\bibnamefont{Alves}},
  \bibinfo{author}{\bibfnamefont{D.}~\bibnamefont{Gon\c{c}alves}},
  \bibinfo{author}{\bibfnamefont{T.}~\bibnamefont{Ghosh}},
  \bibinfo{author}{\bibfnamefont{H.-K.} \bibnamefont{Guo}}, \bibnamefont{and}
  \bibinfo{author}{\bibfnamefont{K.}~\bibnamefont{Sinha}},
  \bibinfo{journal}{JHEP} \textbf{\bibinfo{volume}{03}}, \bibinfo{pages}{053}
  (\bibinfo{year}{2020}), \eprint{1909.05268}.

\bibitem[{\citenamefont{Alves et~al.}(2021)\citenamefont{Alves, Gon\c{c}alves,
  Ghosh, Guo, and Sinha}}]{Alves:2020bpi}
\bibinfo{author}{\bibfnamefont{A.}~\bibnamefont{Alves}},
  \bibinfo{author}{\bibfnamefont{D.}~\bibnamefont{Gon\c{c}alves}},
  \bibinfo{author}{\bibfnamefont{T.}~\bibnamefont{Ghosh}},
  \bibinfo{author}{\bibfnamefont{H.-K.} \bibnamefont{Guo}}, \bibnamefont{and}
  \bibinfo{author}{\bibfnamefont{K.}~\bibnamefont{Sinha}},
  \bibinfo{journal}{Phys. Lett. B} \textbf{\bibinfo{volume}{818}},
  \bibinfo{pages}{136377} (\bibinfo{year}{2021}), \eprint{2007.15654}.

\bibitem[{\citenamefont{Robens and Stefaniak}(2015)}]{Robens:2015gla}
\bibinfo{author}{\bibfnamefont{T.}~\bibnamefont{Robens}} \bibnamefont{and}
  \bibinfo{author}{\bibfnamefont{T.}~\bibnamefont{Stefaniak}},
  \bibinfo{journal}{Eur. Phys. J. C} \textbf{\bibinfo{volume}{75}},
  \bibinfo{pages}{104} (\bibinfo{year}{2015}), \eprint{1501.02234}.

\bibitem[{\citenamefont{Robens and Stefaniak}(2016)}]{Robens:2016xkb}
\bibinfo{author}{\bibfnamefont{T.}~\bibnamefont{Robens}} \bibnamefont{and}
  \bibinfo{author}{\bibfnamefont{T.}~\bibnamefont{Stefaniak}},
  \bibinfo{journal}{Eur. Phys. J. C} \textbf{\bibinfo{volume}{76}},
  \bibinfo{pages}{268} (\bibinfo{year}{2016}), \eprint{1601.07880}.

\bibitem[{\citenamefont{Gonderinger et~al.}(2010)\citenamefont{Gonderinger, Li,
  Patel, and Ramsey-Musolf}}]{Gonderinger:2009jp}
\bibinfo{author}{\bibfnamefont{M.}~\bibnamefont{Gonderinger}},
  \bibinfo{author}{\bibfnamefont{Y.}~\bibnamefont{Li}},
  \bibinfo{author}{\bibfnamefont{H.}~\bibnamefont{Patel}}, \bibnamefont{and}
  \bibinfo{author}{\bibfnamefont{M.~J.} \bibnamefont{Ramsey-Musolf}},
  \bibinfo{journal}{JHEP} \textbf{\bibinfo{volume}{01}}, \bibinfo{pages}{053}
  (\bibinfo{year}{2010}), \eprint{0910.3167}.

\bibitem[{\citenamefont{de~Florian et~al.}(2017)\citenamefont{de~Florian,
  Grojean, Maltoni, Mariotti, Nikitenko, Pieri, Savard, Schumacher, Tanaka,
  Aggleton et~al.}}]{deFlorian:2227475}
\bibinfo{author}{\bibfnamefont{D.}~\bibnamefont{de~Florian}},
  \bibinfo{author}{\bibfnamefont{C.}~\bibnamefont{Grojean}},
  \bibinfo{author}{\bibfnamefont{F.}~\bibnamefont{Maltoni}},
  \bibinfo{author}{\bibfnamefont{C.}~\bibnamefont{Mariotti}},
  \bibinfo{author}{\bibfnamefont{A.}~\bibnamefont{Nikitenko}},
  \bibinfo{author}{\bibfnamefont{M.}~\bibnamefont{Pieri}},
  \bibinfo{author}{\bibfnamefont{P.}~\bibnamefont{Savard}},
  \bibinfo{author}{\bibfnamefont{M.}~\bibnamefont{Schumacher}},
  \bibinfo{author}{\bibfnamefont{R.}~\bibnamefont{Tanaka}},
  \bibinfo{author}{\bibfnamefont{R.}~\bibnamefont{Aggleton}},
  \bibnamefont{et~al.}
  (\bibinfo{collaboration}{LHCHiggsCrossSectionWorkingGroup}),
  \emph{\bibinfo{title}{{Handbook of LHC Higgs Cross Sections: 4. Deciphering
  the Nature of the Higgs Sector}}}, CERN Yellow Reports: Monographs
  (\bibinfo{publisher}{CERN}, \bibinfo{address}{Geneva}, \bibinfo{year}{2017}),
  \bibinfo{note}{869 pages, 295 figures, 248 tables and 1645 citations. Working
  Group web page: https://twiki.cern.ch/twiki/bin/view/LHCPhysics/LHCHXSWG},
  \urlprefix\url{https://cds.cern.ch/record/2227475}.

\bibitem[{\citenamefont{Aaboud et~al.}(2018{\natexlab{d}})}]{ATLAS:2018hxb}
\bibinfo{author}{\bibfnamefont{M.}~\bibnamefont{Aaboud}} \bibnamefont{et~al.}
  (\bibinfo{collaboration}{ATLAS}), \bibinfo{journal}{Phys. Rev. D}
  \textbf{\bibinfo{volume}{98}}, \bibinfo{pages}{052005}
  (\bibinfo{year}{2018}{\natexlab{d}}), \eprint{1802.04146}.

\bibitem[{\citenamefont{Aaboud et~al.}(2019{\natexlab{d}})}]{ATLAS:2018ynr}
\bibinfo{author}{\bibfnamefont{M.}~\bibnamefont{Aaboud}} \bibnamefont{et~al.}
  (\bibinfo{collaboration}{ATLAS}), \bibinfo{journal}{Phys. Rev. D}
  \textbf{\bibinfo{volume}{99}}, \bibinfo{pages}{072001}
  (\bibinfo{year}{2019}{\natexlab{d}}), \eprint{1811.08856}.

\bibitem[{\citenamefont{Aaboud et~al.}(2019{\natexlab{e}})}]{ATLAS:2018xbv}
\bibinfo{author}{\bibfnamefont{M.}~\bibnamefont{Aaboud}} \bibnamefont{et~al.}
  (\bibinfo{collaboration}{ATLAS}), \bibinfo{journal}{Phys. Lett. B}
  \textbf{\bibinfo{volume}{789}}, \bibinfo{pages}{508}
  (\bibinfo{year}{2019}{\natexlab{e}}), \eprint{1808.09054}.

\bibitem[{\citenamefont{Aaboud et~al.}(2018{\natexlab{e}})}]{ATLAS:2017azn}
\bibinfo{author}{\bibfnamefont{M.}~\bibnamefont{Aaboud}} \bibnamefont{et~al.}
  (\bibinfo{collaboration}{ATLAS}), \bibinfo{journal}{JHEP}
  \textbf{\bibinfo{volume}{03}}, \bibinfo{pages}{095}
  (\bibinfo{year}{2018}{\natexlab{e}}), \eprint{1712.02304}.

\bibitem[{\citenamefont{Aaboud et~al.}(2018{\natexlab{f}})}]{ATLAS:2018kot}
\bibinfo{author}{\bibfnamefont{M.}~\bibnamefont{Aaboud}} \bibnamefont{et~al.}
  (\bibinfo{collaboration}{ATLAS}), \bibinfo{journal}{Phys. Lett. B}
  \textbf{\bibinfo{volume}{786}}, \bibinfo{pages}{59}
  (\bibinfo{year}{2018}{\natexlab{f}}), \eprint{1808.08238}.

\bibitem[{\citenamefont{Aaboud et~al.}(2018{\natexlab{g}})}]{ATLAS:2018mme}
\bibinfo{author}{\bibfnamefont{M.}~\bibnamefont{Aaboud}} \bibnamefont{et~al.}
  (\bibinfo{collaboration}{ATLAS}), \bibinfo{journal}{Phys. Lett. B}
  \textbf{\bibinfo{volume}{784}}, \bibinfo{pages}{173}
  (\bibinfo{year}{2018}{\natexlab{g}}), \eprint{1806.00425}.

\bibitem[{\citenamefont{Sirunyan et~al.}(2018{\natexlab{d}})}]{CMS:2018amk}
\bibinfo{author}{\bibfnamefont{A.~M.} \bibnamefont{Sirunyan}}
  \bibnamefont{et~al.} (\bibinfo{collaboration}{CMS}), \bibinfo{journal}{JHEP}
  \textbf{\bibinfo{volume}{06}}, \bibinfo{pages}{127}
  (\bibinfo{year}{2018}{\natexlab{d}}), \bibinfo{note}{[Erratum: JHEP 03, 128
  (2019)]}, \eprint{1804.01939}.

\bibitem[{\citenamefont{Aaboud et~al.}(2018{\natexlab{h}})}]{ATLAS:2018sbw}
\bibinfo{author}{\bibfnamefont{M.}~\bibnamefont{Aaboud}} \bibnamefont{et~al.}
  (\bibinfo{collaboration}{ATLAS}), \bibinfo{journal}{Phys. Rev. D}
  \textbf{\bibinfo{volume}{98}}, \bibinfo{pages}{052008}
  (\bibinfo{year}{2018}{\natexlab{h}}), \eprint{1808.02380}.

\bibitem[{\citenamefont{Aad et~al.}(2016{\natexlab{a}})}]{ATLAS:2015pre}
\bibinfo{author}{\bibfnamefont{G.}~\bibnamefont{Aad}} \bibnamefont{et~al.}
  (\bibinfo{collaboration}{ATLAS}), \bibinfo{journal}{Eur. Phys. J. C}
  \textbf{\bibinfo{volume}{76}}, \bibinfo{pages}{45}
  (\bibinfo{year}{2016}{\natexlab{a}}), \eprint{1507.05930}.

\bibitem[{\citenamefont{Aad et~al.}(2016{\natexlab{b}})}]{ATLAS:2015oxt}
\bibinfo{author}{\bibfnamefont{G.}~\bibnamefont{Aad}} \bibnamefont{et~al.}
  (\bibinfo{collaboration}{ATLAS}), \bibinfo{journal}{Phys. Lett. B}
  \textbf{\bibinfo{volume}{755}}, \bibinfo{pages}{285}
  (\bibinfo{year}{2016}{\natexlab{b}}), \eprint{1512.05099}.

\bibitem[{\citenamefont{Chatrchyan et~al.}(2013)}]{CMS:2013vyt}
\bibinfo{author}{\bibfnamefont{S.}~\bibnamefont{Chatrchyan}}
  \bibnamefont{et~al.} (\bibinfo{collaboration}{CMS}), \bibinfo{journal}{Eur.
  Phys. J. C} \textbf{\bibinfo{volume}{73}}, \bibinfo{pages}{2469}
  (\bibinfo{year}{2013}), \eprint{1304.0213}.

\bibitem[{\citenamefont{Khachatryan et~al.}(2015)}]{CMS:2015hra}
\bibinfo{author}{\bibfnamefont{V.}~\bibnamefont{Khachatryan}}
  \bibnamefont{et~al.} (\bibinfo{collaboration}{CMS}), \bibinfo{journal}{JHEP}
  \textbf{\bibinfo{volume}{10}}, \bibinfo{pages}{144} (\bibinfo{year}{2015}),
  \eprint{1504.00936}.

\bibitem[{\citenamefont{Hagiwara et~al.}(1994)\citenamefont{Hagiwara,
  Matsumoto, Haidt, and Kim}}]{Hagiwara:1994pw}
\bibinfo{author}{\bibfnamefont{K.}~\bibnamefont{Hagiwara}},
  \bibinfo{author}{\bibfnamefont{S.}~\bibnamefont{Matsumoto}},
  \bibinfo{author}{\bibfnamefont{D.}~\bibnamefont{Haidt}}, \bibnamefont{and}
  \bibinfo{author}{\bibfnamefont{C.~S.} \bibnamefont{Kim}},
  \bibinfo{journal}{Z. Phys. C} \textbf{\bibinfo{volume}{64}},
  \bibinfo{pages}{559} (\bibinfo{year}{1994}), \bibinfo{note}{[Erratum:
  Z.Phys.C 68, 352 (1995)]}, \eprint{hep-ph/9409380}.

\bibitem[{\citenamefont{Baak et~al.}(2014)\citenamefont{Baak, C\'uth, Haller,
  Hoecker, Kogler, M\"onig, Schott, and Stelzer}}]{Baak:2014ora}
\bibinfo{author}{\bibfnamefont{M.}~\bibnamefont{Baak}},
  \bibinfo{author}{\bibfnamefont{J.}~\bibnamefont{C\'uth}},
  \bibinfo{author}{\bibfnamefont{J.}~\bibnamefont{Haller}},
  \bibinfo{author}{\bibfnamefont{A.}~\bibnamefont{Hoecker}},
  \bibinfo{author}{\bibfnamefont{R.}~\bibnamefont{Kogler}},
  \bibinfo{author}{\bibfnamefont{K.}~\bibnamefont{M\"onig}},
  \bibinfo{author}{\bibfnamefont{M.}~\bibnamefont{Schott}}, \bibnamefont{and}
  \bibinfo{author}{\bibfnamefont{J.}~\bibnamefont{Stelzer}}
  (\bibinfo{collaboration}{Gfitter Group}), \bibinfo{journal}{Eur. Phys. J. C}
  \textbf{\bibinfo{volume}{74}}, \bibinfo{pages}{3046} (\bibinfo{year}{2014}),
  \eprint{1407.3792}.

\bibitem[{\citenamefont{Quiros}(1994)}]{Quiros:1994dr}
\bibinfo{author}{\bibfnamefont{M.}~\bibnamefont{Quiros}},
  \bibinfo{journal}{Helv. Phys. Acta} \textbf{\bibinfo{volume}{67}},
  \bibinfo{pages}{451} (\bibinfo{year}{1994}).

\bibitem[{\citenamefont{Profumo et~al.}(2015)\citenamefont{Profumo,
  Ramsey-Musolf, Wainwright, and Winslow}}]{Profumo:2014opa}
\bibinfo{author}{\bibfnamefont{S.}~\bibnamefont{Profumo}},
  \bibinfo{author}{\bibfnamefont{M.~J.} \bibnamefont{Ramsey-Musolf}},
  \bibinfo{author}{\bibfnamefont{C.~L.} \bibnamefont{Wainwright}},
  \bibnamefont{and} \bibinfo{author}{\bibfnamefont{P.}~\bibnamefont{Winslow}},
  \bibinfo{journal}{Phys. Rev. D} \textbf{\bibinfo{volume}{91}},
  \bibinfo{pages}{035018} (\bibinfo{year}{2015}), \eprint{1407.5342}.

\bibitem[{\citenamefont{Pietroni}(1993)}]{Pietroni:1992in}
\bibinfo{author}{\bibfnamefont{M.}~\bibnamefont{Pietroni}},
  \bibinfo{journal}{Nucl. Phys. B} \textbf{\bibinfo{volume}{402}},
  \bibinfo{pages}{27} (\bibinfo{year}{1993}), \eprint{hep-ph/9207227}.

\bibitem[{\citenamefont{Morrissey and Ramsey-Musolf}(2012)}]{Morrissey:2012db}
\bibinfo{author}{\bibfnamefont{D.~E.} \bibnamefont{Morrissey}}
  \bibnamefont{and} \bibinfo{author}{\bibfnamefont{M.~J.}
  \bibnamefont{Ramsey-Musolf}}, \bibinfo{journal}{New J. Phys.}
  \textbf{\bibinfo{volume}{14}}, \bibinfo{pages}{125003}
  (\bibinfo{year}{2012}), \eprint{1206.2942}.

\bibitem[{\citenamefont{Wainwright}(2012)}]{Wainwright:2011kj}
\bibinfo{author}{\bibfnamefont{C.~L.} \bibnamefont{Wainwright}},
  \bibinfo{journal}{Comput. Phys. Commun.} \textbf{\bibinfo{volume}{183}},
  \bibinfo{pages}{2006} (\bibinfo{year}{2012}), \eprint{1109.4189}.

\bibitem[{\citenamefont{Alwall et~al.}(2014{\natexlab{a}})\citenamefont{Alwall,
  Frederix, Frixione, Hirschi, Maltoni, Mattelaer, Shao, Stelzer, Torrielli,
  and Zaro}}]{Alwall:2014hca}
\bibinfo{author}{\bibfnamefont{J.}~\bibnamefont{Alwall}},
  \bibinfo{author}{\bibfnamefont{R.}~\bibnamefont{Frederix}},
  \bibinfo{author}{\bibfnamefont{S.}~\bibnamefont{Frixione}},
  \bibinfo{author}{\bibfnamefont{V.}~\bibnamefont{Hirschi}},
  \bibinfo{author}{\bibfnamefont{F.}~\bibnamefont{Maltoni}},
  \bibinfo{author}{\bibfnamefont{O.}~\bibnamefont{Mattelaer}},
  \bibinfo{author}{\bibfnamefont{H.~S.} \bibnamefont{Shao}},
  \bibinfo{author}{\bibfnamefont{T.}~\bibnamefont{Stelzer}},
  \bibinfo{author}{\bibfnamefont{P.}~\bibnamefont{Torrielli}},
  \bibnamefont{and} \bibinfo{author}{\bibfnamefont{M.}~\bibnamefont{Zaro}},
  \bibinfo{journal}{JHEP} \textbf{\bibinfo{volume}{07}}, \bibinfo{pages}{079}
  (\bibinfo{year}{2014}{\natexlab{a}}), \eprint{1405.0301}.

\bibitem[{\citenamefont{Sj\"ostrand et~al.}(2015)\citenamefont{Sj\"ostrand,
  Ask, Christiansen, Corke, Desai, Ilten, Mrenna, Prestel, Rasmussen, and
  Skands}}]{Sjostrand:2014zea}
\bibinfo{author}{\bibfnamefont{T.}~\bibnamefont{Sj\"ostrand}},
  \bibinfo{author}{\bibfnamefont{S.}~\bibnamefont{Ask}},
  \bibinfo{author}{\bibfnamefont{J.~R.} \bibnamefont{Christiansen}},
  \bibinfo{author}{\bibfnamefont{R.}~\bibnamefont{Corke}},
  \bibinfo{author}{\bibfnamefont{N.}~\bibnamefont{Desai}},
  \bibinfo{author}{\bibfnamefont{P.}~\bibnamefont{Ilten}},
  \bibinfo{author}{\bibfnamefont{S.}~\bibnamefont{Mrenna}},
  \bibinfo{author}{\bibfnamefont{S.}~\bibnamefont{Prestel}},
  \bibinfo{author}{\bibfnamefont{C.~O.} \bibnamefont{Rasmussen}},
  \bibnamefont{and} \bibinfo{author}{\bibfnamefont{P.~Z.}
  \bibnamefont{Skands}}, \bibinfo{journal}{Comput. Phys. Commun.}
  \textbf{\bibinfo{volume}{191}}, \bibinfo{pages}{159} (\bibinfo{year}{2015}),
  \eprint{1410.3012}.

\bibitem[{\citenamefont{de~Favereau et~al.}(2014)\citenamefont{de~Favereau,
  Delaere, Demin, Giammanco, Lema\^\i{}tre, Mertens, and
  Selvaggi}}]{deFavereau:2013fsa}
\bibinfo{author}{\bibfnamefont{J.}~\bibnamefont{de~Favereau}},
  \bibinfo{author}{\bibfnamefont{C.}~\bibnamefont{Delaere}},
  \bibinfo{author}{\bibfnamefont{P.}~\bibnamefont{Demin}},
  \bibinfo{author}{\bibfnamefont{A.}~\bibnamefont{Giammanco}},
  \bibinfo{author}{\bibfnamefont{V.}~\bibnamefont{Lema\^\i{}tre}},
  \bibinfo{author}{\bibfnamefont{A.}~\bibnamefont{Mertens}}, \bibnamefont{and}
  \bibinfo{author}{\bibfnamefont{M.}~\bibnamefont{Selvaggi}}
  (\bibinfo{collaboration}{DELPHES 3}), \bibinfo{journal}{JHEP}
  \textbf{\bibinfo{volume}{02}}, \bibinfo{pages}{057} (\bibinfo{year}{2014}),
  \eprint{1307.6346}.

\bibitem[{CMS(2016)}]{CMS-PAS-BTV-15-001}
\bibinfo{type}{Tech. Rep.}, \bibinfo{institution}{(CMS)},
  \bibinfo{address}{Geneva} (\bibinfo{year}{2016}),
  \urlprefix\url{https://cds.cern.ch/record/2138504}.

\bibitem[{\citenamefont{Friedman}(2001)}]{10.1214/aos/1013203451}
\bibinfo{author}{\bibfnamefont{J.~H.} \bibnamefont{Friedman}},
  \bibinfo{journal}{The Annals of Statistics} \textbf{\bibinfo{volume}{29}},
  \bibinfo{pages}{1189 } (\bibinfo{year}{2001}).

\bibitem[{\citenamefont{Hocker et~al.}(2007{\natexlab{a}})}]{Hocker:2007ht}
\bibinfo{author}{\bibfnamefont{A.}~\bibnamefont{Hocker}} \bibnamefont{et~al.}
  (\bibinfo{year}{2007}{\natexlab{a}}), \eprint{physics/0703039}.

\bibitem[{\citenamefont{Alwall et~al.}(2014{\natexlab{b}})\citenamefont{Alwall,
  Frederix, Frixione, Hirschi, Maltoni, Mattelaer, Shao, Stelzer, Torrielli,
  and Zaro}}]{Alwall_2014}
\bibinfo{author}{\bibfnamefont{J.}~\bibnamefont{Alwall}},
  \bibinfo{author}{\bibfnamefont{R.}~\bibnamefont{Frederix}},
  \bibinfo{author}{\bibfnamefont{S.}~\bibnamefont{Frixione}},
  \bibinfo{author}{\bibfnamefont{V.}~\bibnamefont{Hirschi}},
  \bibinfo{author}{\bibfnamefont{F.}~\bibnamefont{Maltoni}},
  \bibinfo{author}{\bibfnamefont{O.}~\bibnamefont{Mattelaer}},
  \bibinfo{author}{\bibfnamefont{H.-S.} \bibnamefont{Shao}},
  \bibinfo{author}{\bibfnamefont{T.}~\bibnamefont{Stelzer}},
  \bibinfo{author}{\bibfnamefont{P.}~\bibnamefont{Torrielli}},
  \bibnamefont{and} \bibinfo{author}{\bibfnamefont{M.}~\bibnamefont{Zaro}},
  \bibinfo{journal}{Journal of High Energy Physics}
  \textbf{\bibinfo{volume}{2014}} (\bibinfo{year}{2014}{\natexlab{b}}),
  \urlprefix\url{https://doi.org/10.1007%2Fjhep07%282014%29079}.

\bibitem[{ttb()}]{ttbarxsec}
\emph{\bibinfo{title}{$\rm{NNLO} + \rm{NNLL}$ top-quark-pair cross sections}},
  \bibinfo{howpublished}{\url{https://twiki.cern.ch/twiki/bin/view/LHCPhysics/TtbarNNLO}}.

\bibitem[{\citenamefont{Goodfellow et~al.}(2016)\citenamefont{Goodfellow,
  Bengio, and Courville}}]{Goodfellow-et-al-2016}
\bibinfo{author}{\bibfnamefont{I.}~\bibnamefont{Goodfellow}},
  \bibinfo{author}{\bibfnamefont{Y.}~\bibnamefont{Bengio}}, \bibnamefont{and}
  \bibinfo{author}{\bibfnamefont{A.}~\bibnamefont{Courville}},
  \emph{\bibinfo{title}{Deep Learning}} (\bibinfo{publisher}{MIT Press},
  \bibinfo{year}{2016}), \bibinfo{note}{\url{http://www.deeplearningbook.org}}.

\bibitem[{\citenamefont{Zech}(1989)}]{Zech:1988un}
\bibinfo{author}{\bibfnamefont{G.}~\bibnamefont{Zech}}, \bibinfo{journal}{Nucl.
  Instrum. Meth. A} \textbf{\bibinfo{volume}{277}}, \bibinfo{pages}{608}
  (\bibinfo{year}{1989}).

\bibitem[{CMS(2018{\natexlab{b}})}]{CMS-PAS-FTR-18-019}
\bibinfo{type}{Tech. Rep.}, \bibinfo{institution}{(CMS)},
  \bibinfo{address}{Geneva} (\bibinfo{year}{2018}{\natexlab{b}}),
  \urlprefix\url{https://cds.cern.ch/record/2652549}.

\bibitem[{\citenamefont{Cowan}(2013)}]{Cowan:2013pha}
\bibinfo{author}{\bibfnamefont{G.}~\bibnamefont{Cowan}}, in
  \emph{\bibinfo{booktitle}{{69th Scottish Universities Summer School in
  Physics}: {LHC Physics}}} (\bibinfo{year}{2013}), pp.
  \bibinfo{pages}{321--355}, \eprint{1307.2487}.

\bibitem[{\citenamefont{Cowan et~al.}(2011)\citenamefont{Cowan, Cranmer, Gross,
  and Vitells}}]{Cowan:2010js}
\bibinfo{author}{\bibfnamefont{G.}~\bibnamefont{Cowan}},
  \bibinfo{author}{\bibfnamefont{K.}~\bibnamefont{Cranmer}},
  \bibinfo{author}{\bibfnamefont{E.}~\bibnamefont{Gross}}, \bibnamefont{and}
  \bibinfo{author}{\bibfnamefont{O.}~\bibnamefont{Vitells}},
  \bibinfo{journal}{Eur. Phys. J. C} \textbf{\bibinfo{volume}{71}},
  \bibinfo{pages}{1554} (\bibinfo{year}{2011}), \bibinfo{note}{[Erratum:
  Eur.Phys.J.C 73, 2501 (2013)]}, \eprint{1007.1727}.

\bibitem[{\citenamefont{Hocker et~al.}(2007{\natexlab{b}})}]{TMVA:2007ngy}
\bibinfo{author}{\bibfnamefont{A.}~\bibnamefont{Hocker}} \bibnamefont{et~al.}
  (\bibinfo{collaboration}{TMVA}) (\bibinfo{year}{2007}{\natexlab{b}}),
  \eprint{physics/0703039}.

\bibitem[{\citenamefont{Freund and Schapire}(1997)}]{freund1997decision}
\bibinfo{author}{\bibfnamefont{Y.}~\bibnamefont{Freund}} \bibnamefont{and}
  \bibinfo{author}{\bibfnamefont{R.~E.} \bibnamefont{Schapire}},
  \bibinfo{journal}{Journal of computer and system sciences}
  \textbf{\bibinfo{volume}{55}}, \bibinfo{pages}{119} (\bibinfo{year}{1997}).

\bibitem[{\citenamefont{Chollet et~al.}(2015)}]{chollet2015keras}
\bibinfo{author}{\bibfnamefont{F.}~\bibnamefont{Chollet}} \bibnamefont{et~al.},
  \emph{\bibinfo{title}{Keras}},
  \bibinfo{howpublished}{\url{https://github.com/keras-team/keras}}
  (\bibinfo{year}{2015}).

\bibitem[{\citenamefont{Pedregosa et~al.}(2011)\citenamefont{Pedregosa,
  Varoquaux, Gramfort, Michel, Thirion, Grisel, Blondel, Prettenhofer, Weiss,
  Dubourg et~al.}}]{scikit-learn}
\bibinfo{author}{\bibfnamefont{F.}~\bibnamefont{Pedregosa}},
  \bibinfo{author}{\bibfnamefont{G.}~\bibnamefont{Varoquaux}},
  \bibinfo{author}{\bibfnamefont{A.}~\bibnamefont{Gramfort}},
  \bibinfo{author}{\bibfnamefont{V.}~\bibnamefont{Michel}},
  \bibinfo{author}{\bibfnamefont{B.}~\bibnamefont{Thirion}},
  \bibinfo{author}{\bibfnamefont{O.}~\bibnamefont{Grisel}},
  \bibinfo{author}{\bibfnamefont{M.}~\bibnamefont{Blondel}},
  \bibinfo{author}{\bibfnamefont{P.}~\bibnamefont{Prettenhofer}},
  \bibinfo{author}{\bibfnamefont{R.}~\bibnamefont{Weiss}},
  \bibinfo{author}{\bibfnamefont{V.}~\bibnamefont{Dubourg}},
  \bibnamefont{et~al.}, \bibinfo{journal}{Journal of Machine Learning Research}
  \textbf{\bibinfo{volume}{12}}, \bibinfo{pages}{2825} (\bibinfo{year}{2011}).

\bibitem[{\citenamefont{Nair and Hinton}(2010)}]{NairHinton2010}
\bibinfo{author}{\bibfnamefont{V.}~\bibnamefont{Nair}} \bibnamefont{and}
  \bibinfo{author}{\bibfnamefont{G.~E.} \bibnamefont{Hinton}},
  \bibinfo{journal}{Proceedings of the 27th International Conference on Machine
  Learning (ICML-10)}  (\bibinfo{year}{2010}).

\bibitem[{\citenamefont{Bridle}(1990)}]{Bridle1990}
\bibinfo{author}{\bibfnamefont{J.~S.} \bibnamefont{Bridle}},
  \bibinfo{journal}{Proceedings of the IEEE International Conference on Neural
  Networks}  (\bibinfo{year}{1990}).

\bibitem[{\citenamefont{Kingma and Ba}(2014)}]{KingmaBa2014}
\bibinfo{author}{\bibfnamefont{D.~P.} \bibnamefont{Kingma}} \bibnamefont{and}
  \bibinfo{author}{\bibfnamefont{J.}~\bibnamefont{Ba}}, \bibinfo{journal}{arXiv
  preprint arXiv:1412.6980}  (\bibinfo{year}{2014}).

\bibitem[{\citenamefont{Rubinstein and Kroese}(2004)}]{RubinsteinKroese2004}
\bibinfo{author}{\bibfnamefont{R.~Y.} \bibnamefont{Rubinstein}}
  \bibnamefont{and} \bibinfo{author}{\bibfnamefont{D.~P.}
  \bibnamefont{Kroese}}, \bibinfo{journal}{Springer}  (\bibinfo{year}{2004}).

\bibitem[{\citenamefont{Sokolova and Lapalme}(2009)}]{SokolovaLapalme2009}
\bibinfo{author}{\bibfnamefont{M.}~\bibnamefont{Sokolova}} \bibnamefont{and}
  \bibinfo{author}{\bibfnamefont{G.}~\bibnamefont{Lapalme}},
  \bibinfo{journal}{Information Processing and Management}
  \textbf{\bibinfo{volume}{45}}, \bibinfo{pages}{427} (\bibinfo{year}{2009}).

\end{thebibliography}
    \end{document}